# A non-relativistic theory of quantum mechanics with local modulus symmetry

Tao Zhou

Department of Physics, New Jersey Institute of Technology, Newark, NJ 07102

## Abstract

We have constructed a non-relativistic theory of quantum mechanics based on local modulus symmetry. The resulting connection in the covariant derivative is identified as the escape velocity of the gravitational field. A new real and positive function called the quantum metric function is attached to the complex conjugate of the wave function to satisfy the local modulus symmetry requirement. In an expanding universe, these theoretical features produce new effects that deviate from predictions of conventional quantum mechanics and Newtonian gravity. The quantum metric function yields negligible change for microscopic objects but produces quantum pointer states for macroscopic objects, thus providing a solution to the quantum measurement problem. The time-reversal symmetry is broken in the new quantum kinematics, which has implications for the second law of thermodynamics. The modification of Newtonian gravity is negligible in smaller bound systems but can become significant at the galactic scales. Its potential association with the mass discrepancy problem in the galactic systems is discussed.

**Email: taozhou@njit.edu**

## I. Introduction

### A. Universal constants

The Planck constant $\hbar$, the speed of light in vacuum $c$, and the gravitational constant $G$ are considered the most fundamental constants in physics. Together, they form the natural unit system, where physical quantities in the Planck scale, *e.g.,* Planck length and Planck time, take the numerical value of 1. One can also examine the fundamental theories of physics from the perspective of these three constants, as shown in Fig. 1. In the standard unit SI system based on human scale, $G$, $\hbar$, and $c^{-1}$ are all quite small numerically, thus the crudest approximation in SI system is to equate all three to 0. This results in, crudely speaking, the "$0^{th}$ order" theory, *i.e.,* classical mechanics, with the kinematics of Newtonian absolute space and time, but without the effect of gravity, as shown on the lowest level in Fig. 1. The next level, "1st order" approximation is to make $c^{-1}$, $G$, and $\hbar$ finite separately.

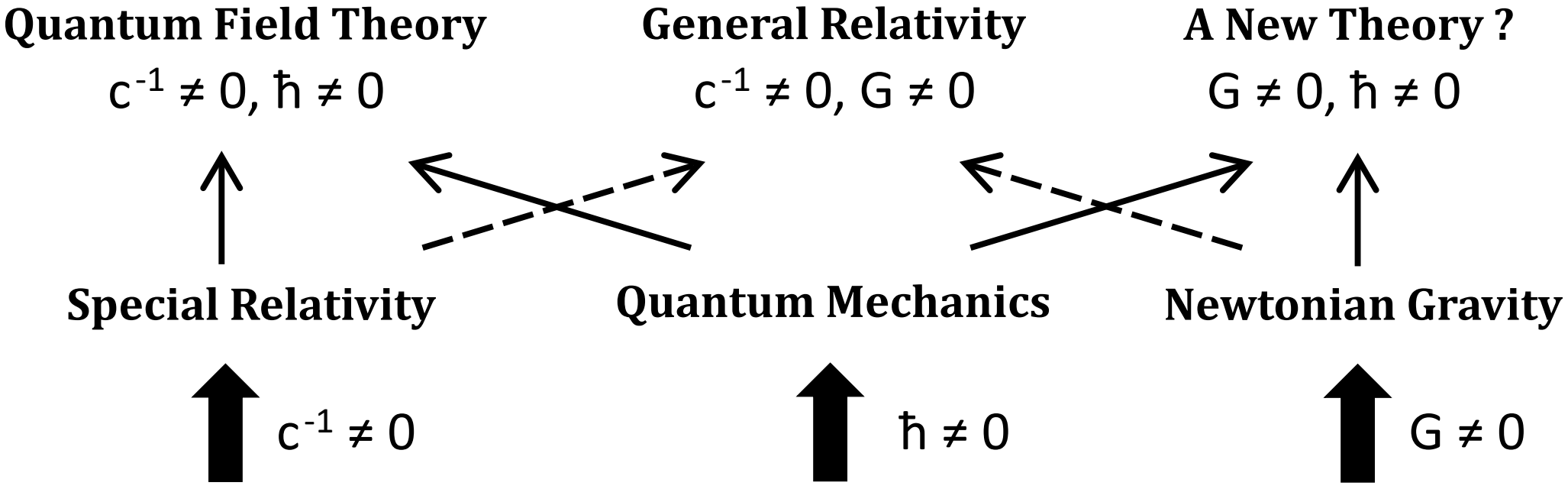

*Figure 1: Fundamental theories in physics perceived from the three fundamental constants: the Planck constant ħ, the speed of light in vacuum c, and the gravitational constant G.*

This results in three theories: special relativity (SR), Newtonian gravity, and non-relativistic quantum mechanics (QM). The next level "2nd order" approximation is to make $c^{-1}$, $G$, and $\hbar$ finite two at a time. Thus, as shown at the highest level in Fig. 1, the theory with finite $c^{-1}$ and $G$ but with $\hbar$ set to 0 is general relativity (GR), the theory with finite $c^{-1}$ and $\hbar$ but with $G$ set to 0 is quantum field theory (QFT). Currently, these two theories are generally considered the most fundamental theories in physics. However, based on the perspective presented in Fig. 1, there should be a third "2nd order" theory, with finite $G$ and $\hbar$ but with $c^{-1}$ set to 0. Sitting at the same level as GR and QFT, this theory is conceivably of fundamental importance as well. Yet, curiously, such a theory has received little attention so far.

Let us consider the simplest situation where both quantum mechanics and gravity should be considered: a quantum particle with mass $m$ inside a gravitation field generated by some external source with mass density $\rho_{ext}(\boldsymbol{r})$. Thus, one has two equations:

$$i\hbar\frac{\partial\psi(\boldsymbol{r},t)}{\partial t}=-\frac{\hbar^2}{2m}\nabla^2\psi(\boldsymbol{r},t)+mu_g(\boldsymbol{r})\psi(\boldsymbol{r},t) \tag{1}$$

$$\nabla^2 u_g(\boldsymbol{r})=4\pi G\rho_{ext}(\boldsymbol{r}) \tag{2}$$

where Eq. (1) is the Schrödinger equation with unit mass gravitational potential energy $u_g(\boldsymbol{r})$ and Eq. (2) is the gravitational Poisson equation.

Penrose *et al.* studied the case where a quantum particle moves in a Newtonian gravitational potential created by the particle's own mass density distribution $m|\psi(\boldsymbol{r},t)|^2$.[1] Replacing $\rho_{ext}(\boldsymbol{r})$ in Eq. (2) by $m|\psi(\boldsymbol{r},t)|^2$, Eq. (1) and (2), now called the Schrödinger-Newton equations by Penrose, become coupled as wave function $\psi(\boldsymbol{r},t)$ appear in both equations. The primary motivation of Penrose *et al.* studying these coupled equations is to tackle the quantum measurement problem by exploring the possibility of gravitation-induced quantum state reduction.[2] It should be noted that similar schemes have been explored before.[3,4] Starting from the generally covariant Newton-Cartan theory of gravity, Christian has used a variational approach to combine it with non-relativistic quantum mechanics, and he also reached the same Schrödinger-Newton equations.[5] Instead of quantizing GR or "general relativizing QFT", he argued that one could try to "special relativize" these Schrödinger-Newton equations, and this approach then constitutes a third way to explore quantum gravity and related Planck scale physics.

Regardless of the motivations, from the perspective of Fig. 1, the approach shown in Eq. (1) and (2), whether coupled or not, remain unsatisfying, mainly in two aspects. First, one of the signatures in any "2nd order" theory is a term or terms containing two universal constants. In the case of GR, they are terms containing both $G$ and $c^{-1}$; in QFT, terms containing both $\hbar$ and $c^{-1}$. Yet in Eq. (1) and (2), no terms explicitly have both $G$ and $\hbar$. Secondly, and more importantly, both GR and QFT are local theories with global symmetries (Lorentz and gauge symmetries, respectively) elevated to local symmetries. The approach of Eq. (1) and (2) does not have this crucial element. One of the main motivations of this paper is to construct a non-relativistic quantum mechanical theory that under local symmetry requirement naturally gives rise to gravity, and in the process can remedy these two deficiencies.

### B. Local symmetries

In order to combine quantum mechanics with gravity, it is helpful to learn from GR, the last successful theory that combines a "1st order" fundamental theory, *i.e.,* special relativity, with gravity. When introducing the new kinematics of SR, Einstein stressed the importance of symmetry and universal constant (in this case, *c*), and this is manifested by the two famous postulates in SR.[6] In the process of developing GR, Einstein recognized the importance of the weak equivalence principle of Newtonian gravity on the gravitational side and the importance of Lorentz symmetry on the SR side. His way of combining them is to elevate the Lorentz symmetry from a global one in SR to a local one in GR, in the form of the Einstein equivalence principle; at the same time, he introduced a new element, the varying metric tensor, into the kinematics to express the effect of gravity.[7]

It is worth emphasizing here the difference between the weak and the Einstein equivalence principles.[8] The weak equivalence principle is essentially the equality between the inertial mass and the gravitational mass of any massive object. It is a firm feature of Newtonian mechanics and gravity, strongly supported by experimental evidence going back to Galileo's time.[9] The Einstein equivalence principle states that in a small enough region of spacetime, the law of non-gravitational physics reduces to those of SR. It is a generalization of the weak equivalence principle in the context of relativistic spacetime.[8] We note that the geometric nature of GR is closely related to the Einstein equivalence principle, but the root of geometrization is not in the weak equivalence principle. Instead, it should be traced back to Minkowski's geometric formulation of SR[10]. In a broader context not specifically tied to relativistic theory, one can argue that the essence of the equivalence principle is the expression of gravity through kinematics instead of dynamics, as gravity is not an ordinary force associated with conventional dynamics, and it can be transformed away locally by certain kinematic transformation.

Learning from QFT, the other "2nd order" fundamental theory is also beneficial. Similar to general covariance in GR, the most crucial symmetry in QFT is the gauge symmetry. One starts from the observation that in QM, different from classical wave, the phase of the wave function has a global symmetry, *i.e.,* one can add an arbitrary uniform phase to the quantum wave function with no observable consequence. This global symmetry is then elevated into a local gauge symmetry, with phase transformation varying from point to point over spacetime. Similar to GR, this also leads to covariant derivatives, and the associated connections are related to gauge fields that correspond to electromagnetic, weak, and strong forces. Interestingly, the wave function in QM has another global symmetry in its modulus. Again, unlike classical waves, one can multiply a constant real and positive number to the unnormalized quantum wave function, and that does not alter anything in the quantum system.

With these insights from the two existing "2nd level" theories, we can start to outline the framework of the theory that combines gravity with QM but without relativity. On the gravitational side, it is natural to take the equivalence principle as the foundation. The equivalence principle here is not the Einstein equivalence principle but should encompass the weak equivalence principle. Additionally, some new elements need to be introduced into the kinematics to express the gravitational effect, similar to the metric tensor in GR. One should also be able to transform away this new element locally so that in small enough regions, the laws of physics reduce to those of conventional QM, just as in GR, they reduce to SR locally. This can serve as the new definition of the equivalence principle in the theory here. On the QM side, we will show that the relevant symmetry is related to the modulus of the wave function. Similar to the global phase symmetry being elevated into a local phase symmetry in QFT, the global symmetry of the modulus of wave function in QM mentioned above will be elevated into a local symmetry in the theory here. The resulting covariant derivative will be shown to give rise to the gravitational field.

## II. Theory

### A. Quantum state

As mentioned in the introduction, due to the similarities of "2nd order theories", the construction of our current theory will be guided in many places by drawing analogies with GR and QFT. In this section, we will highlight the similarity between wave function, its position basis in conventional QM and vector field, and its orthonormal basis in SR. This will set the stage for the discussion in the next section where the position basis of QM needs to be modified under the local symmetry requirement, just like the orthonormal basis in SR is changed into the coordinate basis in GR under the requirement of general covariance.

Mathematically, QM is built on the notion that the state of a quantum system is described by a single vector in the Hilbert space. This state vector can be denoted by a ket vector $|\psi\rangle$, or equally well by its Hermitian conjugate, the bra vector $\langle\psi|$. For reasons that will become clear later, spatial position representation will be used exclusively in the discussion below. Suppose the position basis ket is denoted as $|\boldsymbol{r}\rangle$, and $\langle\boldsymbol{r}|$ is the position basis bra, with $\boldsymbol{r}$ representing any point in space, and suppose $\psi(\boldsymbol{r}) = \langle\boldsymbol{r}|\psi\rangle$, then inserting the identity operator $\int d^3r\, |\boldsymbol{r}\rangle\langle\boldsymbol{r}|$ into $|\psi\rangle$, one has $|\psi\rangle = \int d^3r\, |\boldsymbol{r}\rangle\langle\boldsymbol{r}|\psi\rangle = \int d^3r\, \psi(\boldsymbol{r})|\boldsymbol{r}\rangle$. Similarly, the bra vector $\langle\psi|$ can be decomposed as $\langle\psi| = \int d^3r\, \psi^*(\boldsymbol{r})\langle\boldsymbol{r}|$, where $\psi^*(\boldsymbol{r})$ is the complex conjugate of $\psi(\boldsymbol{r})$. In QM, it is usually taken for granted that the wave function $\psi(\boldsymbol{r})$, or its time-dependent form $\psi(\boldsymbol{r}, t)$, is entirely equivalent to the state vector $|\psi\rangle$, and $\psi^*(\boldsymbol{r},t)$ equivalent to $\langle\psi|$. However, this equivalence is implicitly based on $\int d^3r\,|\boldsymbol{r}\rangle\langle\boldsymbol{r}|$ being an identity operator, which is in turn based on the orthonormal condition:

$$\langle\boldsymbol{r}'|\boldsymbol{r}\rangle = \delta(\boldsymbol{r}'-\boldsymbol{r}) = \delta(x'-x)\delta(y'-y)\delta(z'-z) \qquad (3)$$

Here and in the discussion below, for convenience, we adopt the Cartesian coordinates as the default coordinate system, with the basis ket $|\boldsymbol{r}\rangle = |x,y,z\rangle = |x\rangle|y\rangle|z\rangle$, and the basis bra $\langle\boldsymbol{r}| = \langle z,y,x| = \langle z|\langle y|\langle x|$. $\delta(\boldsymbol{r}'-\boldsymbol{r})$ is the 3-dimensional Dirac delta function.

In SR, any vector $V$, its component denoted as $V^\mu$, has a corresponding dual vector (or covector) $\tilde{V}$, with the component $\tilde{V}_\nu$, and the relation $V^\mu = \eta^{\mu\nu}\tilde{V}_\nu$, where $\eta^{\mu\nu}$ is the Minkowski metric tensor. In QM, the bra vector $\langle\psi|$ was initially introduced by Dirac as the dual vector of the ket vector $|\psi\rangle$ in the Hilbert space.[11] Since the bra vector is the Hermitian conjugate of the ket vector, *i.e.*, $\langle\psi| = |\psi\rangle^\dagger$, $\langle\psi|$ and $|\psi\rangle$ essentially contain the same information, similar to the vector and its dual vector in SR.

To extend the analogy further, we note that in SR, the vector component $V^\mu$ and its dual vector component $\tilde{V}_\nu$, are real fields over spacetime. They can thus be more adequately denoted as $V^\mu(x)$ and $\tilde{V}_\nu(x)$, with $x$ representing any point in spacetime in the context of SR and GR. (In the context of QM, $x$ represents one of the three dimensions of Cartesian coordinates, but since the discussions here are clearly delineated, there should be no confusion.) Since $\eta^{\mu\nu}$ is constant over the whole spacetime, $\tilde{V}_\nu(x)$ contains the same information as $V^\mu(x)$. In QM, the closer analogies to $V^\mu(x)$ and $\tilde{V}_\nu(x)$ are the wave function $\psi(\boldsymbol{r},t)$ and its complex conjugate $\psi^*(\boldsymbol{r},t)$, instead of $|\psi\rangle$ and $\langle\psi|$. This is because $\psi(\boldsymbol{r},t)$ and $\psi^*(\boldsymbol{r},t)$, similar to $V^\mu(x)$ and $\tilde{V}_\nu(x)$, are also functions of space and time, while $|\psi\rangle$ and $\langle\psi|$ are single vectors in the Hilbert space. Let us assume that $\hat{e}_\mu$ and $\hat{e}^{\nu}$ represent the sets of basis vectors and basis dual vectors for the vector field $V(x)$ and dual vector field $\tilde{V}(x)$ in SR, respectively, so $V(x) = V^{\mu}(x)\hat{e}_{\mu}$, and $\tilde{V}(x) = V_{\nu}(x)\hat{e}^{\nu}$. $\psi(\boldsymbol{r},t)$ and $\psi^{*}(\boldsymbol{r},t)$ in QM are thus similar to $V^{\mu}(x)$ and $V_{\nu}(x)$ in SR, while the orthonormal $|\boldsymbol{r}\rangle$ basis ket and $\langle\boldsymbol{r}|$ basis bra in QM play roles similar to the Cartesian-like orthonormal bases of $\hat{e}_{\mu}$ and $\hat{e}^{\nu}$ in SR, *i.e.,* constant over the whole spacetime. When gravity is present and spacetime is curved, a global Cartesian-like coordinate system is not possible anymore. To comply with general covariance, coordinate basis vectors, with varying magnitudes from point to point, are commonly used in GR instead of the orthonormal basis. As shown below, similar effects will be found for the position basis kets and bras in QM when the local symmetry requirement is imposed.

### B. Modulus transformation

We now introduce the local modulus transformation. Similar to the simplest gauge transformation in QFT, *i.e.,* the local phase transformation with *U(1)* symmetry in quantum electrodynamics (QED), one can change the modulus of the non-relativistic wave function of any quantum system by multiplying an addition factor $e^{\lambda(\boldsymbol{r})}$,

$$\psi(\boldsymbol{r},t) \to e^{\lambda(\boldsymbol{r})}\psi(\boldsymbol{r},t) \tag{4}$$

Here $\lambda(\boldsymbol{r})$ is an arbitrary, real, smooth, and differentiable scalar function of spatial position $\boldsymbol{r}$. The exponential form is chosen here to guarantee the factor positive definite.

Note that $\lambda(\boldsymbol{r})$ is only a function of space, and time is not explicitly present. This does not mean one performs this operation continuously over time. Instead, in the context of non-relativistic theory, this operation should be understood as a one-off event. It can be performed at any given time, thus no explicit time in $\lambda(\boldsymbol{r})$, but once the wave function's modulus is changed by the additional factor of $e^{\lambda(\boldsymbol{r})}$, the system is allowed to evolve without further interference. This is different from the QED case because of the non-relativistic nature of the theory here.

We now introduce one of the most essential postulates in our theory here. Similar to the general covariance in GR and the gauge symmetry in QFT, we postulate that the quantum system remains invariant when its wave function undergoes any local modulus transformation.

Before we discuss the consequences of this postulate, it is worthwhile to note the lessons in GR again. In GR, one may take the vector field $V(x)$ and covector field $\tilde{V}(x)$ as abstract geometric objects over spacetime, and express them as $V(x) = V^{\mu}(x)\hat{e}_{\mu}(x)$ and $\tilde{V}(x) = V_{\nu}(x)\hat{e}^{\nu}(x)$, with $\hat{e}_{\mu}(x)$ and $\hat{e}^{\nu}(x)$ now representing the coordinate basis vectors and dual vectors that vary over spacetime. The components $V^{\mu}(x)$ and $V_{\nu}(x)$ transform inversely as their respective basis vectors and basis dual vectors $\hat{e}_{\mu}(x)$ and $\hat{e}^{\nu}(x)$ do, indicated by their respective super and subscript indices. This guarantees $V(x)$ and $\tilde{V}(x)$ as geometric invariant objects under a local coordinate transformation.

In the previous section, we draw the analogy between $V^{\mu}(x)\hat{e}_{\mu}$ in SR and $\psi(\boldsymbol{r},t)|\boldsymbol{r}\rangle$ in QM. Let us formalize it with the definition of $\acute{\psi}(\boldsymbol{r},t)$ as

$$\acute{\psi}(\boldsymbol{r},t) = \psi(\boldsymbol{r},t)|\boldsymbol{r}\rangle \tag{5}$$

and we can extend the analogy between $V(x)$ in GR and $\acute{\psi}(\boldsymbol{r},t)$ in the theory here by identifying $\acute{\psi}(\boldsymbol{r},t)$ as an object that stays invariant under local modulus transformation. Since $\psi(\boldsymbol{r},t)$ transforms according to Eq. (4), to keep $\acute{\psi}(\boldsymbol{r},t)$ invariant, similar to the coordinate basis in GR, we demand $|\boldsymbol{r}\rangle$ to transform inversely as:

$$|\boldsymbol{r}\rangle \rightarrow e^{-\lambda(r)}|\boldsymbol{r}\rangle \tag{6}$$

On the other hand, since the Dirac delta function is an invariant, the orthonormal condition between $|\boldsymbol{r}\rangle$ and $\langle\boldsymbol{r}|$ in Eq. (3) means $\langle\boldsymbol{r}|$ should transform as:

$$\langle\boldsymbol{r}| \rightarrow e^{\lambda(r)}\langle\boldsymbol{r}| \tag{7}$$

To complete the analogy between vector and covector here and in GR, we define another invariant object $\grave{\psi}(\boldsymbol{r},t)$ as:

$$\grave{\psi}(\boldsymbol{r},t) = \tilde{\psi}(\boldsymbol{r},t)\langle\boldsymbol{r}| \tag{8}$$

$\tilde{\psi}(\boldsymbol{r},t)$ is a new complex scalar function of space and time. Since $\langle\boldsymbol{r}|$ transforms the same way as $\psi(\boldsymbol{r},t)$, to keep $\grave{\psi}(\boldsymbol{r},t)$ invariant, $\tilde{\psi}(\boldsymbol{r},t)$ should transform inversely as $\psi(\boldsymbol{r},t)$:

$$\tilde{\psi}(\boldsymbol{r},t) \rightarrow e^{-\lambda(r)}\tilde{\psi}(\boldsymbol{r},t) \tag{9}$$

We note that the covector of conventional QM is identified as $\psi^*(\boldsymbol{r},t)\langle\boldsymbol{r}|$ in the previous discussion. $\psi^*(\boldsymbol{r},t)$, being the complex conjugate of $\psi(\boldsymbol{r},t)$, shares with $\psi(\boldsymbol{r},t)$ the same modulus. Thus $\psi^*(\boldsymbol{r},t)$ transforms the same way as $\psi(\boldsymbol{r},t)$ under local modulus transformation and cannot form an invariant object with $\langle\boldsymbol{r}|$. This is why we need a new complex scalar function $\tilde{\psi}(\boldsymbol{r},t)$ in Eq. (8). However, since the phase of any complex function is not altered by the modulus transformation, $\tilde{\psi}(\boldsymbol{r},t)$ is defined to share the same phase as $\psi^*(\boldsymbol{r},t)$. Alternatively, we can define a new real positive dimensionless smooth function $\gamma(\boldsymbol{r})$ as the ratio between the modulus of $\tilde{\psi}(\boldsymbol{r},t)$ and $\psi^*(\boldsymbol{r},t)$ so that

$$\tilde{\psi}(\boldsymbol{r},t) = \gamma(\boldsymbol{r})\psi^*(\boldsymbol{r},t) \tag{10}$$

with a transformation rule of

$$\gamma(\boldsymbol{r}) \rightarrow e^{-2\lambda(r)}\gamma(\boldsymbol{r}) \tag{11}$$

to guarantee the transformation rule of $\tilde{\psi}(\boldsymbol{r},t)$ as Eq. (9). We call $\tilde{\psi}(\boldsymbol{r},t)$ the co-wave function.

With the introduction of $\acute{\psi}(\boldsymbol{r},t)$, $\grave{\psi}(\boldsymbol{r},t)$ $\tilde{\psi}(\boldsymbol{r},t)$ and $\gamma(\boldsymbol{r})$, as well as the transformation rules for $\psi(\boldsymbol{r},t)$, $\tilde{\psi}(\boldsymbol{r},t)$, $|\boldsymbol{r}\rangle$ and $\langle\boldsymbol{r}|$, we can now investigate the consequences of this postulate, first on the kinematics of QM in the next section, then on the dynamics of QM in the section after.

### C. Quantum kinematics

In conventional QM, all observables are real numbers that result from some combination of $\psi(\boldsymbol{r},t)$ and $\psi^*(\boldsymbol{r},t)$. To keep these observables invariant to local modulus transformation, one needs to replace $\psi^*(\boldsymbol{r},t)$ by $\tilde{\psi}(\boldsymbol{r},t)$. For example, the probability density $\rho$ is now defined as $\rho(\boldsymbol{r},t) = \tilde{\psi}(\boldsymbol{r},t)\,\psi(\boldsymbol{r},t)$, which is invariant due to the inversely related transformation rules of $\psi(\boldsymbol{r},t)$ and $\tilde{\psi}(\boldsymbol{r},t)$. To transition from QM to the theory here, the first thing to do is to systematically go over everything and replace $\psi^*(\boldsymbol{r},t)$ by $\tilde{\psi}(\boldsymbol{r},t)$, or equivalently by $\gamma(\boldsymbol{r})\psi^*(\boldsymbol{r},t)$.

The next object whose transformation property needs to be considered is the spatial derivative of the wave function $\boldsymbol{\nabla}\psi(\boldsymbol{r},t)$. In order to make it behave under local modulus transformation, the mathematical treatment is very similar to the local phase transformation in QED. We first define the covariant spatial derivative $\boldsymbol{D_r}$ that will be used to replace the ordinary derivative $\boldsymbol{\nabla}$ as $\boldsymbol{D_r} = \boldsymbol{\nabla} + \boldsymbol{Y}(\boldsymbol{r})$, with $\boldsymbol{Y}(\boldsymbol{r})$ being the vector connection. Thus

$$\boldsymbol{D_r}\psi(\boldsymbol{r},t) = \boldsymbol{\nabla}\psi(\boldsymbol{r},t) + \boldsymbol{Y}(\boldsymbol{r})\psi(\boldsymbol{r},t) \tag{12}$$

Under the local modulus transformation $\boldsymbol{Y}(\boldsymbol{r})$ should transform as:

$$\boldsymbol{Y}(\boldsymbol{r}) \to \boldsymbol{Y}(\boldsymbol{r}) - \boldsymbol{\nabla}\lambda(\boldsymbol{r}) \tag{13}$$

so that we have

$$\boldsymbol{D_r}\psi(\boldsymbol{r}) \rightarrow [\boldsymbol{\nabla} + \boldsymbol{Y}(\boldsymbol{r}) - \boldsymbol{\nabla}\lambda(\boldsymbol{r})]\left[e^{\lambda(r)}\psi(\boldsymbol{r})\right] = e^{\lambda(r)}[\boldsymbol{D_r}\psi(\boldsymbol{r})] \tag{14}$$

Thus, the covariant spatial derivative of $\psi(\boldsymbol{r})$ transforms exactly like $\psi(\boldsymbol{r})$ itself. To conform with local modulus symmetry, the second thing to do is to systematically replace the ordinary derivative $\boldsymbol{\nabla}$ by the covariant derivative $\boldsymbol{D_r}$.

Unlike conventional QM, the position basis ket $|\boldsymbol{r}\rangle$ and bra $\langle\boldsymbol{r}|$ can change from place to place here, implied by the transformation rules in Eq. (6) and (7). Thus, one expects the spatial derivatives of $|\boldsymbol{r}\rangle$ and $\langle\boldsymbol{r}|$ to play essential roles in our theory. However, we will not discuss them in this section or in the next section. This is due to the similarity between the basis ket and bra we use here and the coordinate basis in GR. Because of the convenient transformation rules for tensors and their bases in the coordinate basis system, calculations in GR can be performed almost exclusively on tensor components and their covariant derivatives with their associated connection, without worrying about the varying basis vectors and their derivatives. In discussions on the kinematics and dynamics of quantum particles, a similar strategy adopted for $|\boldsymbol{r}\rangle$ and $\langle\boldsymbol{r}|$ here will allow us to focus only on $\psi(\boldsymbol{r}, t)$ and $\tilde{\psi}(\boldsymbol{r}, t)$, with gravitational effect confined in the connection $\boldsymbol{Y}(\boldsymbol{r})$ alone. Later on, when we discuss how to relate $\boldsymbol{Y}(\boldsymbol{r})$ with $|\boldsymbol{r}\rangle$ and $\langle\boldsymbol{r}|$, we then need to discuss the details of $|\boldsymbol{r}\rangle$, $\langle\boldsymbol{r}|$, and their derivatives.

We now move on to quantum operators. As we discussed before, with position representation, the operator $\int d^3r\, |\boldsymbol{r}\rangle\langle\boldsymbol{r}|$ is an identity operator due to the orthonormal condition in Eq. (3). Since this condition remains the same even when $|\boldsymbol{r}\rangle$ and $\langle\boldsymbol{r}|$ change from place to place, $\int d^3r\, |\boldsymbol{r}\rangle\langle\boldsymbol{r}|$ stays as an identity operator, and the projector operator $P_r = |\boldsymbol{r}\rangle\langle\boldsymbol{r}|$ also remains invariant. Essentially, both the inner and outer product of $|\boldsymbol{r}\rangle$ and $\langle\boldsymbol{r}|$ remain invariant, as their transformation rules in Eq. (6) and (7) dictate the same results

in both cases. Consequently, we do not expect any change in the form of quantum operators in position representation here. This is also one of the reasons why position representation is used exclusively in this paper.

We can rewrite the kinematics of quantum particles based on the 3 principles laid out above: replacing $\psi^*(\boldsymbol{r},t)$ by $\tilde{\psi}(\boldsymbol{r},t)$, replacing $\boldsymbol{\nabla}$ by $\boldsymbol{D_r}$, and using the same form of quantum operators. For example, the expectation value of particle position should be $\langle\hat{\boldsymbol{r}}\rangle = \int d^3r\, \tilde{\psi}(\boldsymbol{r},t)\boldsymbol{r}\psi(\boldsymbol{r},t)$, where $\hat{\boldsymbol{r}}$ is the position operator.

Let us construct the particle momentum operator $\hat{\boldsymbol{p}}$ based on the covariant derivative:

$$\hat{\boldsymbol{p}} = \frac{\hbar}{i}\boldsymbol{D_r} = \frac{\hbar}{i}\boldsymbol{\nabla} + \frac{\hbar}{i}\boldsymbol{Y}(\boldsymbol{r}) = \frac{\hbar}{i}\boldsymbol{\nabla} - im\boldsymbol{v^e}(\boldsymbol{r}) \tag{15}$$

Here we have replaced the connection $\boldsymbol{Y}(\boldsymbol{r})$ by a velocity field $\boldsymbol{v^e}(\boldsymbol{r})$ with

$$\boldsymbol{v^e}(\boldsymbol{r}) = \frac{\hbar}{m}\,\boldsymbol{Y}(\boldsymbol{r}) \tag{16}$$

and $m$ is the mass of the particle.

The Hermitian conjugate of $\hat{\boldsymbol{p}}$, denoted as $\hat{\boldsymbol{p}}^\dagger$, is easily shown as:

$$\hat{\boldsymbol{p}}^\dagger = \hat{\boldsymbol{p}}^{*T} = \frac{\hbar}{i}\boldsymbol{\nabla} - \frac{\hbar}{i}\boldsymbol{Y}(\boldsymbol{r}) = \frac{\hbar}{i}\boldsymbol{\nabla} + im\boldsymbol{v^e}(\boldsymbol{r}) \tag{17}$$

thus $\hat{\boldsymbol{p}}^\dagger \neq \hat{\boldsymbol{p}}$. This raises the question of whether momentum-related physical quantities in our theory have real eigenvalues. Since Hermitian operators are closely related to the unitarity of the quantum system, this also raises the even more serious question of whether the quantum state can evolve unitarily in our theory. We will address the first question now, and the second in the next section, with positive answers in both cases.

We first note that the definition of covariant derivative of $\psi(\boldsymbol{r},t)$ in Eq. (12) and its transformation rule in Eq. (14) cannot be applied to $\tilde{\psi}(\boldsymbol{r},t)$. Instead, we need to define a different covariant derivative $\widetilde{\boldsymbol{D}}_{\boldsymbol{r}}$ for $\tilde{\psi}(\boldsymbol{r},t)$ as:

$$\widetilde{\boldsymbol{D}}_{\boldsymbol{r}}\tilde{\psi}(\boldsymbol{r},t) = \boldsymbol{\nabla}\tilde{\psi}(\boldsymbol{r},t) - \boldsymbol{Y}(\boldsymbol{r})\tilde{\psi}(\boldsymbol{r},t) \tag{18}$$

with the same connection $\boldsymbol{Y}(\boldsymbol{r})$. Inserting its transformation rule in Eq. (13) one has:

$$\widetilde{\boldsymbol{D}}_{\boldsymbol{r}}\tilde{\psi}(\boldsymbol{r}) \to [\boldsymbol{\nabla} - \boldsymbol{Y}(\boldsymbol{r}) + \boldsymbol{\nabla}\lambda(\boldsymbol{r})]\left[e^{-\lambda(r)}\tilde{\psi}(\boldsymbol{r})\right] = e^{-\lambda(r)}[\widetilde{\boldsymbol{D}}_{\boldsymbol{r}}\tilde{\psi}(\boldsymbol{r})] \tag{19}$$

and $\widehat{\boldsymbol{p}}^{\dagger} = \frac{\hbar}{i}\widetilde{\boldsymbol{D}}_{\boldsymbol{r}}$ compared to $\widehat{\boldsymbol{p}} = \frac{\hbar}{i}\boldsymbol{D}_{\boldsymbol{r}}$. We note that in QED, there is only one definition of the covariant derivative. However, in GR the covariant derivatives for the vector field and covector field are very similar to those in Eq.(12) and (18). It is also obvious here that once $\tilde{\psi}(\boldsymbol{r},t)$ is used instead of $\psi^*(\boldsymbol{r},t)$, it is inevitable that $\widehat{\boldsymbol{p}}^{\dagger} \neq \widehat{\boldsymbol{p}}$.

In conventional QM, the expectation value of $\widehat{\boldsymbol{p}}$ is:

$$\overline{\boldsymbol{p}} = \langle\psi|\widehat{\boldsymbol{p}}\psi\rangle = \int \psi^*(\boldsymbol{r},t)\frac{\hbar}{i}\boldsymbol{\nabla}\psi(\boldsymbol{r},t)d^3r \tag{20}$$

The probability density flow $\boldsymbol{j}(\boldsymbol{r},t)$, on the other hand, is defined locally as:

$$\boldsymbol{j}(\boldsymbol{r},t) = \frac{1}{2m}[\psi^*(\boldsymbol{r},t)\widehat{\boldsymbol{p}}\psi(\boldsymbol{r},t) - \psi(\boldsymbol{r},t)\widehat{\boldsymbol{p}}\psi^*(\boldsymbol{r},t)] = \frac{1}{m}\mathrm{Re}[\psi^*\widehat{\boldsymbol{p}}\psi] \tag{21}$$

where Re stands for the real part. If the modulus of $\psi(\boldsymbol{r},t)$ is $R(\boldsymbol{r},t)$ and its phase is $S(\boldsymbol{r},t)/\hbar$ so that we have $\psi(\boldsymbol{r},t) = R(\boldsymbol{r},t)e^{iS(\boldsymbol{r},t)/\hbar}$, and keep in mind that the particle probability density $\rho(\boldsymbol{r},t) = R^2(\boldsymbol{r},t)$ in conventional QM, one has:

$$\overline{\boldsymbol{p}} = \int \rho(\boldsymbol{r},t)\boldsymbol{\nabla}S(\boldsymbol{r},t)d^3r \tag{22}$$

$$\boldsymbol{j}(\boldsymbol{r},t) = \frac{1}{m}\rho(\boldsymbol{r},t)\boldsymbol{\nabla}S(\boldsymbol{r},t) \tag{23}$$

This means $\overline{\boldsymbol{p}}$, which is real by its definition, and $\boldsymbol{j}(\boldsymbol{r},t)$, which is taken as a real part of a complex quantity $\psi^*\widehat{\boldsymbol{p}}\psi$, are essentially built from the same local quantity $\rho(\boldsymbol{r},t)\boldsymbol{\nabla}S(\boldsymbol{r},t)$ in QM, *i.e.*, the gradient of the phase weighted by the probability density.

We stick to the same principle in our theory here. When we make the redefinition of any physical observable that is related to momentum, this expectation value has to be real, and locally, it should be built solely from $\rho(\boldsymbol{r},t)\boldsymbol{\nabla}S(\boldsymbol{r},t)$, with $\rho(\boldsymbol{r},t)$ redefined as:

$$\rho(\boldsymbol{r},t)=\tilde{\psi}(\boldsymbol{r},t)\psi(r,t)=\gamma(\boldsymbol{r})R^2(\boldsymbol{r},t) \tag{24}$$

We note that local modulus transformation does not affect $S(\boldsymbol{r},t)$, thus $\rho(\boldsymbol{r},t)\boldsymbol{\nabla}S(\boldsymbol{r},t)$ is an invariant to this transformation. For $\boldsymbol{j}(\boldsymbol{r},t)$, we can just replace $\psi^*(\boldsymbol{r},t)$ by $\tilde{\psi}(\boldsymbol{r},t)$, use Eq. (15) for the momentum operator $\widehat{\boldsymbol{p}}$, and otherwise use the same definition in Eq. (21) for $\boldsymbol{j}(\boldsymbol{r},t)$. One can then easily show that this results in the same Eq. (23) for $\boldsymbol{j}(\boldsymbol{r},t)$. For the expectation value of momentum, we redefine it as:

$$\overline{\boldsymbol{p}}=\int Re[\tilde{\psi}(\boldsymbol{r},t)\widehat{\boldsymbol{p}}\psi(\boldsymbol{r},t)]d^3r=\int Re[\tilde{\psi}(\boldsymbol{r},t)\frac{\hbar}{i}\boldsymbol{\nabla}\psi(\boldsymbol{r},t)]d^3r \tag{25}$$

Again, one can easily show that this leads to the same Eq. (22) for $\overline{\boldsymbol{p}}$.

In QED, the canonical momentum operator, defined as $\frac{\hbar}{i}\partial_\mu$, corresponds to the total momentum that includes both parts from the particle and the electromagnetic field. One needs to subtract the 4-potential from it to get the particle momentum. Our theory here has a purely imaginary momentum field $im\boldsymbol{v}^{\boldsymbol{e}}(\boldsymbol{r})$, and it does not play any role in the particle momentum and probability density flow, following the definitions in Eq. (23) and (25). Physically, this is consistent with the notion that the relativistic field has its own momentum, while the non-relativistic field acts instantaneously and does not have its own momentum.

In summary, we have outlined a self-consistent method that allows us to reformulate quantum kinematics under local modulus symmetry requirement systematically. After replacing $\psi^*(\boldsymbol{r},t)$ by $\tilde{\psi}(\boldsymbol{r},t)$, and replacing $\boldsymbol{\nabla}$ by $\boldsymbol{D_r}$, one can use the same form of quantum operators as in conventional QM, and all expectation values are defined either as before (if they are already real), or as the real part of the original form (if they are complex). This procedure guarantees that all expectation values are real, invariant to local modulus transformation, and revert to their QM forms when $\boldsymbol{Y}(\boldsymbol{r}) = 0$.

**D. Quantum dynamics**

So far, we have focused on local modulus symmetry's impact on kinematics, and all the discussions are in the context of occurring at the same given time. Now, we turn our attention to their effects on the time evolution of the wave function and the dynamic equation of motion. It is important to stress that since our theory is non-relativistic, the local modulus transformation here is a one-off event. In other words, the transformation can be done at any given moment, but once it is done, the quantum system evolves by itself without any further interference.

Suppose two adjacent points, P and Q, locate in a one-dimensional space. A wave packet, described by the wave function $\psi(x,t)$, passes through space. Suppose at time t, the peak of the wave packet is at point P, and at time $t+\delta t$, this peak reaches point Q. In the case where the wave packet is long with the energy centered around $E = p^2/2m$, the group velocity is $p/m$, so the displacement from P to Q is $\delta x = \delta t \cdot p/m$. In the case of a general wave packet made of superposition of plane waves with a wide spread of different wavelengths, one needs to replace $p$ by the canonical momentum operator $\hat{P} = \frac{\hbar}{i}\partial_x$. Now, suppose one makes a local modulus transformation at time $t$, which results in an additional factor of $e^{\lambda(x_P)}$ at point P and a different factor $e^{\lambda(x_Q)}$ at Q. Since $\boldsymbol{j}(\boldsymbol{r},t)$ is an invariant, the group velocity remains the same, so at time $t+\delta t$, the peak of the wave packet reaches point Q again.

To find out the time derivative $\partial\psi(x,t)/\partial t$ at point Q, one needs to compare the wave function $\psi(x_Q,t)$ with $\psi(x_Q,t+\delta t)$, the latter originating from the wave function at P and time $t$, *i.e.*, $\psi(x_P,t)$. However, at time $t$, the wave function is modulated differently at P and Q by the local modulus transformation. Thus, $\partial\psi(x,t)/\partial t$ at Q, after the transformation, not only depends on the factor $e^{\lambda(x_Q)}$, but also on $e^{\lambda(x_P)}$. As a result, $\partial\psi(x,t)/\partial t$ does not transform covariantly as $\psi(x,t)$. The same argument can obviously be generalized into the 3-dimensional case.

To make both sides of the Schrödinger equation covariant to local modulus transformation, one has to introduce a new form of time derivative, named here as the covariant time derivative $D_t$, with the property that $D_t\psi(\boldsymbol{r},t)$ transforms the same way as $\psi(\boldsymbol{r},t)$. To obtain the exact form of $D_t$, we use the example above but now generalized to 3 dimensions. Since one knows that the wave function $\psi(\boldsymbol{r}_Q,t+\delta t)$ originates from $\psi(\boldsymbol{r}_P,t)$ and thus transforms the same way as $\psi(\boldsymbol{r}_P,t)$, one can use the connection $\boldsymbol{Y}(\boldsymbol{r})$ to transport $\psi(\boldsymbol{r}_Q,t+\delta t)$, so that the resulting wave function, $\psi(P\rightarrow Q,t+\delta t)$, is assumed to transform the same as $\psi(\boldsymbol{r}_Q,t)$. Based on the same relation of parallel transport and covariant derivative of vectors in GR, Eq. (12) then implies that $\psi(P\rightarrow Q,t+\delta t)=\psi(\boldsymbol{r}_Q,t+\delta t)+\boldsymbol{Y}(\boldsymbol{r}_Q)\cdot(\boldsymbol{r}_Q-\boldsymbol{r}_P)\psi(\boldsymbol{r}_Q,t+\delta t)$. As everything is based on point Q, let us define $\delta\boldsymbol{r}=\boldsymbol{r}_P-\boldsymbol{r}_Q$, so $\boldsymbol{r}_Q$ serves as the origin. To the first order approximation, one can replace $\psi(\boldsymbol{r}_Q, t+\delta t)$ in the last term by $\psi(\boldsymbol{r}_Q, t)$ as $\boldsymbol{r}_Q - \boldsymbol{r}_P = -\delta\boldsymbol{r}$ is already a first order infinitesimal. Putting everything together, one can now define the temporal covariant derivative at point Q as:

$$D_t\psi(\boldsymbol{r}_Q, t) = \frac{\psi(P \to Q, t+\delta t) - \psi(\boldsymbol{r}_Q, t)}{\delta t}$$

$$= \frac{\psi(\boldsymbol{r}_Q, t+\delta t) - \psi(\boldsymbol{r}_Q, t) - \boldsymbol{Y}(\boldsymbol{r}_Q)\cdot\delta\boldsymbol{r}\,\psi(\boldsymbol{r}_Q, t)}{\delta t} \qquad (26)$$

The terms in this definition are now all based on the point Q, and since Q can be any point, we will omit this index below, and Eq. (26) can be written as:

$$D_t\psi(\boldsymbol{r}, t) = \frac{\partial\psi(\boldsymbol{r}, t)}{\partial t} - \boldsymbol{Y}(\boldsymbol{r})\cdot\frac{\delta\boldsymbol{r}}{\delta t}\psi(\boldsymbol{r}, t) \qquad (27)$$

As mentioned above, $\delta\boldsymbol{r} = \delta t\cdot\widehat{\boldsymbol{P}}/m = \delta t\cdot\frac{\hbar}{im}\boldsymbol{\nabla}$, and $\boldsymbol{Y}(\boldsymbol{r}) = \frac{m}{\hbar}\boldsymbol{v}^{\boldsymbol{e}}(\boldsymbol{r})$, thus one can rewrite the above equation as follows:

$$D_{\boldsymbol{t}}\psi(\boldsymbol{r}, t) = \frac{\partial\psi(\boldsymbol{r}, t)}{\partial t} + i\boldsymbol{v}^{\boldsymbol{e}}(\boldsymbol{r})\cdot\boldsymbol{\nabla}\psi(\boldsymbol{r}, t) \qquad (28)$$

Having obtained both the spatial and temporal covariant derivatives that transform the same way as $\psi(\boldsymbol{r}, t)$, we can now rewrite the Schrödinger equation to make it covariant to local modulus transformation as:

$$i\hbar D_{\boldsymbol{t}}\psi(\boldsymbol{r}, t) = H\psi(\boldsymbol{r}, t) = \frac{\widehat{\boldsymbol{p}}^{\boldsymbol{2}}}{2m}\psi(\boldsymbol{r}, t) = \frac{-\hbar^2}{2m}\boldsymbol{D_r}^{\boldsymbol{2}}\psi(\boldsymbol{r}, t) \qquad (29)$$

Here, we assume that the system is not subject to any force other than that related to gravity. One can certainly insert any other real potential energy $U(\boldsymbol{r})$ into the Hamiltonian so $H = \frac{-\hbar^2}{2m}\boldsymbol{D_r}^{\boldsymbol{2}} + U(\boldsymbol{r})$, and it is evident that the equation is still covariant to local modulus transformation. However, since this paper is focused on gravity and quantum mechanics, we will ignore any other type of potential energy's effect in the following discussion.

One can now insert Eq. (15) and (28) into Eq. (29), and after expanding the quadratic $\widehat{\boldsymbol{p}}^{\mathbf{2}}$ term, one has:

$$\left\{i\hbar\frac{\partial}{\partial t}-\hbar\boldsymbol{v}^{\boldsymbol{e}}(\boldsymbol{r})\cdot\boldsymbol{\nabla}\right\}\psi(\boldsymbol{r},t)$$

$$=\left\{\frac{-\hbar^2}{2m}\boldsymbol{\nabla}^{\mathbf{2}}-\hbar\boldsymbol{v}^{e}(\boldsymbol{r})\cdot\boldsymbol{\nabla}-\frac{1}{2}m\boldsymbol{v}^{e}(\boldsymbol{r})^2-\frac{\hbar}{2}\boldsymbol{\nabla}\cdot\boldsymbol{v}^{e}(\boldsymbol{r})\right\}\psi(\boldsymbol{r},t) \quad (30)$$

The second terms on both sides of the equation are the same, canceling it out yields:

$$i\hbar\frac{\partial}{\partial t}\,\psi(\boldsymbol{r},t)=\left\{\frac{-\hbar^2}{2m}\boldsymbol{\nabla}^{\mathbf{2}}-\frac{1}{2}m\boldsymbol{v}^{e}(\boldsymbol{r})^2-\frac{\hbar}{2}\boldsymbol{\nabla}\cdot\boldsymbol{v}^{e}(\boldsymbol{r})\right\}\psi(\boldsymbol{r},t) \quad (31)$$

In Newtonian gravity, if $U_g(\boldsymbol{r})$ is the gravitational potential energy of a particle with a mass $m$ at a given point $\boldsymbol{r}$, one can define the escape velocity $\boldsymbol{v}^{\boldsymbol{e}}$ at this point as $U_g(\boldsymbol{r})+\frac{1}{2}m\boldsymbol{v}^{\boldsymbol{e}}(\boldsymbol{r})^2=0$, or

$$U_g(\boldsymbol{r})=-\frac{1}{2}m\boldsymbol{v}^{e}(\boldsymbol{r})^2 \quad (32)$$

This is precisely the second term on the right side of Eq. (31). We thus identify $\boldsymbol{v}^{\boldsymbol{e}}(\boldsymbol{r})$, defined in Eq. (16), as the escape velocity of the gravitational field. Eq. (31) is, therefore, the quantum equation of motion of a particle in an external gravitational field. There is one new term, $-\frac{\hbar}{2}\boldsymbol{\nabla}\cdot\boldsymbol{v}^{e}(\boldsymbol{r})$, in Eq. (31). Since escape velocity always contains gravitational constant $G$, this term thus has both $\hbar$ and $G$ inside it. Adding the fact that Eq. (31) is covariant to local modulus transformation, we have thus fulfilled the goal of remedying the two deficiencies in Eq. (1), as mentioned in the introduction. With both $\hbar$ and $G$ inside it, this last term in Eq. (31) is expected to be relatively small compared to the second term.

However, it is interesting that this term, as some form of particle potential energy, is independent of particle mass. We will discuss it in detail in the next part.

It is worth noting here that in Newtonian gravity, escape velocity, as defined in Eq. (32), is not an accurate name, since only the magnitude of $\boldsymbol{v}^e(\boldsymbol{r})$ is well-defined. Therefore, it should be called escape speed. The escape velocity $\boldsymbol{v}^e(\boldsymbol{r})$, as defined in Eq. (16), is a vector field linearly proportional to the connection $\boldsymbol{Y}(\boldsymbol{r})$. Thus, it not only has direction but also transforms, similar to $\boldsymbol{Y}(\boldsymbol{r})$, as:

$$\boldsymbol{v}^e(\boldsymbol{r}) \to \boldsymbol{v}^e(\boldsymbol{r}) - \hbar \boldsymbol{\nabla} \lambda(\boldsymbol{r})/m \qquad (33)$$

The exact relation between the gravitational source and $\boldsymbol{v}^e(\boldsymbol{r})$ will be developed in the next two sections.

The definitions of both spatial and temporal covariant derivatives above are based on a velocity field $\boldsymbol{v}^e(\boldsymbol{r})$. It thus bears some similarities to fluid dynamics. Heuristically, one can imagine that a flow of purely imaginary velocity $i\boldsymbol{v}^e(\boldsymbol{r})$ fill the space, and gravitational information is encoded in $\boldsymbol{v}^e(\boldsymbol{r})$. If a quantum particle's momentum relative to the space is described by the canonical momentum operator $\widehat{\boldsymbol{P}} = \frac{\hbar}{i}\boldsymbol{\nabla}$, then this particle's momentum relative to the flow should be expressed as $\frac{\hbar}{i}\boldsymbol{\nabla} - im\boldsymbol{v}^e(\boldsymbol{r})$. This is the same as the definition in Eq. (15), based on the spatial covariant derivative. On the other hand, a particle in the flow of velocity field $i\boldsymbol{v}^e(\boldsymbol{r})$ has a Lagrangian derivative of $\partial_t + i\boldsymbol{v}^e(\boldsymbol{r}) \cdot \boldsymbol{\nabla}$. Thus, it matches the temporal covariant derivative that we obtained in Eq. (28).

Consistent with the non-relativistic theory here, the purely imaginary nature of this velocity field does not contribute anything to the observed particle momentum, as we already discussed before. However, the square of the same momentum contributes a real and negative definite term of $-\frac{1}{2}m\boldsymbol{v}^e(\boldsymbol{r})^2$ to the Hamiltonian. This matches the nature of gravity, which is always attractive and always produces a negative potential energy.

In quantum mechanics, if the potential energy in Eq. (1) is of real value, then the Schrödinger equation guarantees that the dynamic evolution is unitary, as the continuity equation can be derived from Eq. (1). In our current theory, with the new definition of probability density in Eq. (24), and its associated density flow defined in Eq. (23), can the dynamic equation of motion, as shown in Eq. (31), still guarantee the unitary evolution of the system? We will present the proof below that it is indeed the case.

Let us define a reduced Hamiltonian $H_r$ as:

$$H_r = \frac{-\hbar^2}{2m}\boldsymbol{\nabla}^2 - \frac{1}{2}m\boldsymbol{v}^e(\boldsymbol{r})^2 - \frac{\hbar}{2}\boldsymbol{\nabla}\cdot\boldsymbol{v}^e(\boldsymbol{r}) \tag{34}$$

So now we can rewrite the Schrödinger equation in Eq. (31) as:

$$i\hbar\frac{\partial}{\partial t}\,\psi(\boldsymbol{r},t) = H_r\psi(\boldsymbol{r},t) \tag{35}$$

Even though the total Hamiltonian of the system is not Hermitian, $H_r$ in Eq. (34) is Hermitian. Based on the same derivation as in conventional QM, we can have the following continuity equation:

$$\frac{\partial\rho_0}{\partial t} + \boldsymbol{\nabla}\cdot\boldsymbol{j}_0 = 0 \tag{36}$$

Here $\rho_0 = \psi^*(\boldsymbol{r},t)\psi(\boldsymbol{r},t)$, while $\boldsymbol{j}_0 = \frac{1}{m}\mathrm{Re}[\psi^*\hat{\boldsymbol{p}}\psi]$, as defined in conventional QM. On the other hand, the particle probability density $\rho$ and density flow $\boldsymbol{j}$ in our theory are dependent on $\tilde{\psi}(\boldsymbol{r},t)$ instead of $\psi^*(r,t)$. Thus $\rho(\boldsymbol{r},t) = \gamma(\boldsymbol{r})\rho_0(\boldsymbol{r},t)$, and $\boldsymbol{j}(\boldsymbol{r},t) = \gamma(\boldsymbol{r})\boldsymbol{j}_0(\boldsymbol{r},t)$. Note that $\gamma(\boldsymbol{r})$ is not a function of time, so $\partial_t\rho(\boldsymbol{r},t) = \gamma(\boldsymbol{r})\partial_t\rho_0(\boldsymbol{r},t)$. On the other hand, $\boldsymbol{\nabla}\cdot\boldsymbol{j}(\boldsymbol{r},t) = \gamma(\boldsymbol{r})\boldsymbol{\nabla}\cdot\boldsymbol{j}_0(\boldsymbol{r},t) + \boldsymbol{\nabla}\gamma(\boldsymbol{r})\cdot\boldsymbol{j}_0(\boldsymbol{r},t)$. Since we have the freedom of attaching an arbitrary factor of $e^{-2\lambda(\boldsymbol{r})}$ to $\gamma(\boldsymbol{r})$, as shown in Eq. (11), we can set $\boldsymbol{\nabla}\gamma(\boldsymbol{r}) = 0$ at any given point. Thus, $\boldsymbol{\nabla}\cdot\boldsymbol{j}(\boldsymbol{r},t) = \gamma(\boldsymbol{r})\boldsymbol{\nabla}\cdot\boldsymbol{j}_0(\boldsymbol{r},t)$ under such a condition, (or take the QFT language, in such a gauge). Multiply a factor $\gamma(\boldsymbol{r})$ onto Eq. (36), we find

$$\gamma(\boldsymbol{r})\frac{\partial\rho_0}{\partial t} + \gamma(\boldsymbol{r})\boldsymbol{\nabla}\cdot\boldsymbol{j}_0 = \frac{\partial\rho}{\partial t} + \boldsymbol{\nabla}\cdot\boldsymbol{j} = 0 \tag{37}$$

Since both $\rho$ and $\boldsymbol{j}$ are invariant to local modulus transformation, if the continuity equation above is valid in one modulus condition, it should be valid in all conditions. Thus, we prove that the quantum system always evolves unitarily in our theory.

### E. Metric functions

Having found the equation of motion in the last section to replace the conventional Schrödinger equation with gravitational potential in Eq. (1), we now can search for the field equation to replace the Poisson equation of Newtonian gravity in Eq. (2). Again, we look for guidance from GR. In GR, the connection in the equation of motion is the Christoffel symbol, denoted here as $\Gamma^{\lambda}_{\mu\nu}$, while the field equation is about finding the solution for the metric tensor $g_{\mu\nu}$. The mathematical relation between $g_{\mu\nu}$ and $\Gamma^{\lambda}_{\mu\nu}$ is facilitated by the Einstein equivalence principle. In this section, we will first define entities called quantum metric functions, which will play a similar role like $g_{\mu\nu}$ in GR. We then proceed to find out its relationship with the connection $Y(\boldsymbol{r})$. In the next section, we will discuss the field equation that can determine the quantum metric functions.

In previous sections of this theory part, we focused on the wave function, quantum operators, and dynamic equations, but deferred the discussion on the basis ket $|\boldsymbol{r}\rangle$ and basis bra $\langle \boldsymbol{r}|$. To study their relationship with $Y(\boldsymbol{r})$, we can analyze the gradient of $\acute{\psi}(\boldsymbol{r},t)$. Since $\acute{\psi}(\boldsymbol{r},t)$ is defined as an object invariant to local modulus transformation, $\boldsymbol{\nabla}\acute{\psi}(\boldsymbol{r},t)$ should also be an invariant. On the other hand, based on the Leibnitz rule, one has:

$$\boldsymbol{\nabla}\acute{\psi}(\boldsymbol{r},t) = \boldsymbol{\nabla}[\psi(\boldsymbol{r},t)|\boldsymbol{r}\rangle] = [\boldsymbol{\nabla}\psi(\boldsymbol{r},t)]|\boldsymbol{r}\rangle + \psi(\boldsymbol{r},t)[\boldsymbol{\nabla}|\boldsymbol{r}\rangle] \tag{38}$$

Since $\psi(\boldsymbol{r},t)$ and $|\boldsymbol{r}\rangle$ transform inversely, to keep $\boldsymbol{\nabla}\acute{\psi}(\boldsymbol{r},t)$ invariant, we need to set

$$\boldsymbol{\nabla}|\boldsymbol{r}\rangle = \boldsymbol{Y}(\boldsymbol{r})|\boldsymbol{r}\rangle \tag{39}$$

Using the same procedure with $\boldsymbol{\nabla}\grave{\psi}(\boldsymbol{r},t)$, we have

$$\boldsymbol{\nabla}\langle\boldsymbol{r}| = -\boldsymbol{Y}(\boldsymbol{r})\langle\boldsymbol{r}| \tag{40}$$

The above two equations are symbolically quite simple but not very useful for obtaining the exact function of $\boldsymbol{Y}(\boldsymbol{r})$. The reason is that while the gradient $\boldsymbol{\nabla}$ is an operator in real space, $|\boldsymbol{r}\rangle$ and $\langle\boldsymbol{r}|$ are basis vector and covector in the Hilbert space. Again, this is like the situation in GR, where the basis vectors of adjacent points reside in different tangent spaces. Consequently, taking a derivative of them directly is not very meaningful. The way around this in GR is to define the metric tensor $g_{\mu\nu}(x)$ as $g_{\mu\nu}(x) = \hat{e}_\mu(x)\cdot\hat{e}_\upsilon(x)$, where $\hat{e}_\mu(x)$ and $\hat{e}_\upsilon(x)$ are the coordinate basis vectors. The gravitational information encoded in $\hat{e}_\mu(x)$ and $\hat{e}_\upsilon(x)$ are thus transferred to $g_{\mu\nu}(x)$, a rank 2 tensorial function of spacetime position $x$. By relating the derivatives of $g_{\mu\nu}(x)$ to the connection $\Gamma^{\lambda}_{\mu\nu}(x)$ and solving the field equation for $g_{\mu\nu}(x)$, one can then find the exact form for $\Gamma^{\lambda}_{\mu\nu}(x)$.

We note that in GR $g_{\mu\nu}(x)$ is a (0, 2) type tensor. In QM, other than the ket vectors and bra covectors, the physical entities we usually encounter are either scalars resulting from the inner products of kets and bras or quantum operators as the outer products of kets and bras. In GR language, if the ket and bra are (1, 0) and (0, 1) type tensors, respectively, then these quantum operators are (1, 1) mixed rank tensors. Due to the orthonormal conditions between $|\boldsymbol{r}\rangle$ and $\langle\boldsymbol{r}|$, even when $|\boldsymbol{r}\rangle$ and $\langle\boldsymbol{r}|$ are not constant over space, these operators stay invariant to local modulus transformation. To extract gravitational information out of $|\boldsymbol{r}\rangle$ and $\langle\boldsymbol{r}|$, we need to construct type (0, 2) tensors like $g_{\mu\nu}(x)$ instead. Let us first define 3 new operators and 3 related scalar functions as:

$$
\begin{aligned}
\grave{\grave{T}}_1\,|x\rangle &= \gamma_1(\boldsymbol{r})\,\langle x| \\
\grave{\grave{T}}_2|y\rangle &= \gamma_2(\boldsymbol{r})\,\langle y| \\
\grave{\grave{T}}_3|z\rangle &= \gamma_3(\boldsymbol{r})\,\langle z|
\end{aligned} \tag{41}
$$

Due to the orthonormal condition in Eq. (3), Eq. (41) above implies that:

$$
\begin{aligned}
\grave{\grave{T}}_1 &= \gamma_1(\boldsymbol{r})\,\langle x|\langle x| \\
\grave{\grave{T}}_2 &= \gamma_2(\boldsymbol{r})\,\langle y|\langle y| \\
\grave{\grave{T}}_3 &= \gamma_3(\boldsymbol{r})\,\langle z|\langle z|
\end{aligned} \tag{42}
$$

Note here $\langle bra|_i$, with $i = 1, 2, 3$, corresponds to the basis bra $\langle x|$, $\langle y|$, and $\langle z|$, respectively. $\grave{\grave{T}}_i$ in Eq. (42) above can be understood as three (0, 2) tensors in their respective dimensions, with $\langle x|\langle x|$, $\langle y|\langle y|$, and $\langle z|\langle z|$ serving as the tensor bases, while $\gamma_1(\boldsymbol{r})$, $\gamma_2(\boldsymbol{r})$, $\gamma_3(\boldsymbol{r})$ as the tensor components. Note that these basis tensors look like the basis bra of composite systems in conventional QM, but of course, they represent very different things. $\grave{\grave{T}}_i$ are (0, 2) type operators in the Hilbert space, while $\gamma_i(\boldsymbol{r})$ are assumed to be smooth, dimensionless, real, and positive definite functions of $\boldsymbol{r}$, so $\gamma_i(\boldsymbol{r})$ are scalar functions in real space.

We further define the complex conjugate operator $T_0$ as:

$$
T_0\,\psi(\boldsymbol{r}, t) = \psi^*(\boldsymbol{r}, t) \tag{43}
$$

so $T_0$ turns wave function into its complex conjugate. Since ${T_0}^2\psi(\boldsymbol{r},t) = \psi(\boldsymbol{r},t)$, the inverse operator of $T_0$, denoted as ${T_0}^{-1}$, is the same as $T_0$, *i.e.,* $T_0 = {T_0}^{-1}$.

Finally, we define the operator $\grave{\grave{T}}$ as:

$$\grave{\grave{T}} = T_0\grave{\grave{T}}_1\grave{\grave{T}}_2\grave{\grave{T}}_3 \tag{44}$$

with $\grave{\grave{T}}$ having the following property:

$$\grave{\grave{T}}\acute{\psi}(\boldsymbol{r},t) = \grave{\psi}(\boldsymbol{r},t) \tag{45}$$

By combining Eq. (42), (43), (44), (45), and (10), we find that

$$\gamma(\boldsymbol{r}) = \gamma_1(\boldsymbol{r})\gamma_2(\boldsymbol{r})\gamma_3(\boldsymbol{r}) \tag{46}$$

Similar to the role the metric tensor plays in GR of turning vector into their covector, the combined effect of $\grave{\grave{T}}_{1-3}$ with $T_0$ is to convert the vector $\acute{\psi}(\boldsymbol{r},t)$ into its covector $\grave{\psi}(\boldsymbol{r},t)$ in our theory here, as shown in Eq. (45). We thus call $\gamma_1(\boldsymbol{r})$, $\gamma_2(\boldsymbol{r})$ and $\gamma_3(\boldsymbol{r})$ the quantum metric functions.

When the local modulus transformation is performed, how does $\gamma_i(\boldsymbol{r})$ transform? To find that, we need to make an important assumption about how $\langle x|$, $\langle y|$, and $\langle z|$ transform. We postulate that the extra factor $e^{\lambda(r)}$ is always distributed evenly between them, *i.e.,*

$$\langle x| \rightarrow e^{\lambda(r)/3}\langle x|, \qquad \langle y| \rightarrow e^{\lambda(r)/3}\langle y|, \qquad \langle z| \rightarrow e^{\lambda(r)/3}\langle z| \tag{47}$$

Since we assume $\grave{\grave{T}}_i$ are tensor-like objects that are invariant to local modulus transformation, from Eq. (47), we know the transformation properties of $\gamma_i(\boldsymbol{r})$ must be:

$$\gamma_i(\boldsymbol{r}) \rightarrow e^{-\frac{2\lambda(\boldsymbol{r})}{3}}\gamma_i(\boldsymbol{r}) \tag{48}$$

Combining them leads to $\gamma(\boldsymbol{r}) \rightarrow e^{-2\lambda(\boldsymbol{r})}\gamma(\boldsymbol{r})$ in Eq. (11).

We denote the inverse operators of $\grave{\grave{T}}_i$ as $\acute{\acute{T}}_i$, so $\acute{\acute{T}}_i = \grave{\grave{T}}_i^{\,-1}$. They are (2, 0) type tensors in the Hilbert space. Apply them to Eq. (42), one can get their expressions as:

$$\begin{aligned} \acute{\acute{T}}_1 &= \gamma_1(\boldsymbol{r})^{-1}\,|x\rangle|x\rangle \\ \acute{\acute{T}}_2 &= \gamma_2(\boldsymbol{r})^{-1}\,|y\rangle|y\rangle \\ \acute{\acute{T}}_3 &= \gamma_3(\boldsymbol{r})^{-1}\,|z\rangle|z\rangle \end{aligned} \tag{49}$$

Note that all operators defined above commute with each other. Finally, we find the inverse of $\grave{\grave{T}}$ as:

$$\acute{\acute{T}} = \grave{\grave{T}}^{-1} = (T_0\grave{\grave{T}}_1\grave{\grave{T}}_2\grave{\grave{T}}_3)^{-1} = T_0\acute{\acute{T}}_1\acute{\acute{T}}_2\acute{\acute{T}}_3 \tag{50}$$

so

$$\acute{\acute{T}}\grave{\psi}(\boldsymbol{r},t) = \acute{\psi}(\boldsymbol{r},t) \tag{51}$$

As mentioned above, we have defined the basis ket and bra in some way similar to the coordinate basis in GR. This allows us to focus on the components of tensors without concerns about their bases. That was the case in previous sections, and it is also the case here, where we do not really need to deal with operators like $\acute{\acute{T}}_i$ or $\grave{\grave{T}}_i$. Instead, from now on, we can focus our attention only on $\gamma_i(\boldsymbol{r})$, and study how they are related to $\boldsymbol{Y}(\boldsymbol{r})$ and $\boldsymbol{v}^{\boldsymbol{e}}(\boldsymbol{r})$, and how they are decided by the gravitational source.

Like $g_{\mu\nu}$, $\gamma_i(\boldsymbol{r})$ is expected to be determined entirely by gravitational effect. Thus, when gravity can be neglected, $\gamma_1(\boldsymbol{r})$, $\gamma_2(\boldsymbol{r})$, and $\gamma_3(\boldsymbol{r})$ should all be constants over space. In the presence of gravity, as $\gamma_1(\boldsymbol{r})$, $\gamma_2(\boldsymbol{r})$, and $\gamma_3(\boldsymbol{r})$ are all different functions, they change differently over space. $\gamma(\boldsymbol{r})$, based on Eq. (46), is the product of $\gamma_1(\boldsymbol{r})$, $\gamma_2(\boldsymbol{r})$, and $\gamma_3(\boldsymbol{r})$, and it can be changed by an arbitrary factor of $e^{-2\lambda(r)}$ through local modulus transformation. However, according to their transformation rules in Eq. (48), the ratios between $\gamma_1(\boldsymbol{r})$, $\gamma_2(\boldsymbol{r})$, and $\gamma_3(\boldsymbol{r})$ are not changed by this transformation. It is precisely these ratios as a function of $\boldsymbol{r}$ that encode the gravitational information.

Having found the transformation rule for $\gamma_{1-3}(\boldsymbol{r})$ in Eq. (48), we still need to find the form of the covariant derivative for $\gamma_{1-3}(\boldsymbol{r})$, as this will be the next step toward associating $\gamma_i(\boldsymbol{r})$ with the connection $\boldsymbol{Y}(\boldsymbol{r})$. Note that the vector connection $\boldsymbol{Y}(\boldsymbol{r})$ is first proposed to construct the covariant derivative of the scalar wave function $\psi(\boldsymbol{r}, t)$ under local modulus transformation. Now there are 3 kets of $|x\rangle$, $|y\rangle$, and $|z\rangle$ multiplied together that represent the basis ket $|\boldsymbol{r}\rangle$ at any position, not the single value R representing the modulus of the scalar wave function. Therefore, connecting such states for adjacent points in 3 spatial dimensions requires a $3 \times 3$ matrix connection, denoted here as $Y_{\boldsymbol{ij}}(\boldsymbol{r})$, with $i = 1, 2, 3$, corresponding to the 3 basis kets, and $j = x, y, z$, corresponding to the 3 spatial directions. We note that the three spatial derivatives ($\partial_x$, $\partial_y$, and $\partial_z$) of the three $\gamma_i$ also form a $3 \times 3$ matrix $\partial_j \gamma_i$, and so do their covariant derivatives, denoted here as $D_j \gamma_i$. While a vector connection $\boldsymbol{Y}(\boldsymbol{r})$ is enough for the kinematics and dynamics of quantum particles, as shown in previous sections, one now needs the rank 2 tensor field $Y_{ij}(\boldsymbol{r})$ to construct the covariant derivatives of $\gamma_i(\boldsymbol{r})$, *i.e.,* to relate $D_j \gamma_i$ with $\partial_j \gamma_i$.

If we insert $|\boldsymbol{r}\rangle = |x\rangle|y\rangle|z\rangle$ into Eq. (39), together with the definition of

$$\partial_j |x\rangle = Y_{1j}(\boldsymbol{r})|x\rangle\,; \qquad \partial_j |y\rangle = Y_{2j}(\boldsymbol{r})|y\rangle\,; \qquad \partial_j |z\rangle = Y_{3j}(\boldsymbol{r})|z\rangle \tag{52}$$

we can easily prove that $Y_j(\boldsymbol{r})$, the $j$th component of $\boldsymbol{Y}(\boldsymbol{r})$, is linked to $Y_{ij}(\boldsymbol{r})$ as:

$$Y_j(\boldsymbol{r}) = Y_{\boldsymbol{1j}}(\boldsymbol{r}) + Y_{\boldsymbol{2j}}(\boldsymbol{r}) + Y_{\boldsymbol{3j}}(\boldsymbol{r}) \tag{53}$$

so $\boldsymbol{Y}(\boldsymbol{r})$ and $Y_{\boldsymbol{ij}}(\boldsymbol{r})$ are closely linked. However, the fact that we now have two types of connections is a new feature that has no parallels in GR. The origin of this is the different number of components vs. bases in $\acute{\psi}(\boldsymbol{r}, t)$ and $\grave{\psi}(\boldsymbol{r}, t)$. In GR, the number of vector components and the number of basis vectors are always the same. Here, for the ket field $\acute{\psi}(\boldsymbol{r},t)$, we have one scalar component $\psi(\boldsymbol{r},t)$, yet three ket bases $|x\rangle$, $|y\rangle$, and $|z\rangle$ that change differently over space. Furthermore, the product of $|x\rangle$, $|y\rangle$, and $|z\rangle$ has a physical meaning, but not the summation of them. In GR, the opposite is true. These differences make the covariant derivative of $g_{\mu\nu}$ in GR less of a guide for constructing the covariant derivative of $\gamma_i(\boldsymbol{r})$ here. Instead, we need to find solutions by examining $\gamma_i(\boldsymbol{r})$'s own characteristics.

Let us first imagine that we can move the quantum (0, 2) tensor $\grave{\grave{T}}$ from one point to its neighboring point in a way similar to the parallel transport in GR. This can help us find out how the covariant derivatives of the components of $\grave{\grave{T}}$, namely $D_j\gamma_i$, differ from their ordinary derivatives $\partial_j\gamma_i$. We first note that as long as the orthonormal coordinate system is maintained here, any process, including the parallel transport process, cannot change $\gamma_1$ into a mixture of $\gamma_1$ with $\gamma_2$ and/or $\gamma_3$. This kind of mixing is as meaningless as mixing $x$ with $y$ or $z$ coordinate in a single Cartesian coordinate system. This means that the covariant derivative of a specific $\gamma_i$ should not be related to the other two $\gamma_i$. Assume the parallel transport process here, like that in GR, only involves linear change, we can conclude that $D_j\gamma_i - \partial_j\gamma_i$ should be proportional to the specific $\gamma_i$ only, and unrelated to the other two $\gamma_i$. This is quite different from the situation in GR.

Since $\gamma_i$ make up the components of tensor $\grave{\grave{T}}$, the difference between the covariant and ordinary derivative of $\gamma_i$ should stem entirely from the change of the bases of $\grave{\grave{T}}$. Eq. (52) shows that the changes of these bases result into $Y_{ij}(\boldsymbol{r})$. Therefore, we can expect that other than $\gamma_i$, the matrix $D_j\gamma_i - \partial_j\gamma_i$ should be entirely dependent on the matrix $Y_{ij}$. We further note that when $\overset{\hat{\leftrightarrow}}{T}$ is parallel transported, the basis tensor involved is $\langle \boldsymbol{r}|\langle \boldsymbol{r}|$. Since $\overset{\hat{\leftrightarrow}}{T}$ is a product of three $\overset{\hat{\leftrightarrow}}{T}_i$ that commute with each other, inside $\overset{\hat{\leftrightarrow}}{T}$ the associative law of multiplication means one can choose to lump $\gamma_1$ with $\langle y|\langle y|$ as legitimately as to lump $\gamma_1$ with $\langle x|\langle x|$. Consequently, one would expect the covariant derivative of $\gamma_1$, for example, to be not only related to $Y_{1j}(\boldsymbol{r})$, but also related to $Y_{2j}(\boldsymbol{r})$ and $Y_{3j}(\boldsymbol{r})$ as well. Therefore, each component of the matrix $D_j\gamma_i - \partial_j\gamma_i$ is not only related to the $Y_{ij}(\boldsymbol{r})$ component with the same *i* and *j*, but also related to other components inside the connection matrix.

How is the matrix $D_j\gamma_i - \partial_j\gamma_i$ related to the matrix $Y_{ij}(\boldsymbol{r})$ then? We note that the deviation of basis bras $\langle x|$, $\langle y|$, and $\langle z|$ from those in conventional QM can be divided into two categories. One is the difference among $\langle x|$, $\langle y|$, and $\langle z|$ at a given position $\boldsymbol{r}$, and the other is the change of basis bras along the 3 different directions in real space. We postulate that the first type can be accounted for by the $Y_{ij}(\boldsymbol{r})$ components with the same $i$ index (representing the $i$ th bra) averaged over three different $j$ (representing 3 different directions), while the second type by the $Y_{ij}(\boldsymbol{r})$ components with the same $j$ averaged over three different $i$. Putting these discussions together, we postulate that the covariant derivative of $\gamma_i(\boldsymbol{r})$ can be expressed as:

$$D_j\gamma_i = \partial_j\gamma_i - \frac{1}{3}(Y_{ix} + Y_{iy} + Y_{iz})\gamma_i - \frac{1}{3}(Y_{1j} + Y_{2j} + Y_{3j})\gamma_i \qquad (54)$$

Note that the last two terms on the right side of Eq. (54) have negative signs. This is because $\gamma_i$ are components of (0, 2) type tensor $\overset{\hat{\leftrightarrow}}{T}$.

Earlier, we have postulated that the extra factor $e^{\lambda(\boldsymbol{r})}$ is always distributed evenly among $\langle x|$, $\langle y|$, and $\langle z|$. This leads to the transformation rule for $\gamma_i$ in Eq. (48). Applying the local modulus transformation to Eq. (54), together with the transformation rule for $\boldsymbol{Y}(\boldsymbol{r})$ in Eq. (13), and $\boldsymbol{Y}(\boldsymbol{r})$ as the sum of its component $Y_{ij}$ shown in Eq. (53), we find that the matrix $Y_{ij}$ needs to be symmetric, *i.e.* $Y_{ij} = Y_{ji}$, to ensure that $\gamma_i(\boldsymbol{r})$ and $\boldsymbol{Y}(\boldsymbol{r})$ transform consistently. Conversely, we can start with the assumption of a symmetric $Y_{ij}$, and conclude that the local modulus transformation always distributes the extra factor $e^{\lambda(\boldsymbol{r})}$ equally among $\langle x|$, $\langle y|$, and $\langle z|$. Note that the connection in GR, the Christoffel symbol $\Gamma^{\lambda}_{\mu\nu}$, is assumed to be torsion free, therefore, also symmetric in its $\mu$ and $v$ indices.

In the introduction part, we postulate a new equivalence principle. It states that in any small enough region, the laws of physics reduce to those of conventional QM. Since in conventional QM $\gamma_i$ are constant over space, it is natural to demand that the connection here be metric compatible, *i.e.,* $D_j\gamma_i = 0$. Again, this is very similar to the situation in GR. Consequently, we have:

$$\partial_j\gamma_i = \frac{1}{3}\left(Y_{ix} + Y_{iy} + Y_{iz}\right)\gamma_i + \frac{1}{3}\left(Y_{1j} + Y_{2j} + Y_{3j}\right)\gamma_i \tag{55}$$

Since $Y_{ij}$ is symmetric, this equation means:

$$\partial_j \ln \gamma_i(\boldsymbol{r}) = \partial_i \ln \gamma_j(\boldsymbol{r}) \tag{56}$$

Let us now define $v^e_{ij}(\boldsymbol{r}) = \frac{\hbar}{m}Y_{ij}(\boldsymbol{r})$, and a new real-valued function $\tau_i(\boldsymbol{r})$ as:

$$\tau_i(\boldsymbol{r}) = \frac{\hbar}{m} \ln \gamma_i(\boldsymbol{r}) \tag{57}$$

For reasons that will become clear later, we call $\tau_i(\boldsymbol{r})$ the regularized metric functions. We also define the sum of them as $\tau(\boldsymbol{r})$:

$$\tau(\boldsymbol{r}) = \tau_1(\boldsymbol{r}) + \tau_2(\boldsymbol{r}) + \tau_3(\boldsymbol{r}) = \frac{\hbar}{m} \ln \gamma(\boldsymbol{r}) \tag{58}$$

Based on Eq. (11) we find that the transformation rule for $\tau(\boldsymbol{r})$ should be

$$\tau(\boldsymbol{r}) \to \tau(\boldsymbol{r}) - 2\hbar\lambda(\boldsymbol{r})/m \tag{59}$$

Eq. (55) can then be written as:

$$\partial_j \tau_i = \frac{1}{3}(v_{ix}^e + v_{iy}^e + v_{iz}^e) + \frac{1}{3}(v_{1j}^e + v_{2j}^e + v_{3j}^e) \tag{60}$$

Note that based on Eq. (53) we can write the $j$th component of the escape velocity $\boldsymbol{v^e}(\boldsymbol{r})$ as $v_j^e(\boldsymbol{r})$ with $v_j^e(\boldsymbol{r})$ as:

$$v_j^e(\boldsymbol{r}) = v_{1j}^e(\boldsymbol{r}) + v_{2j}^e(\boldsymbol{r}) + v_{3j}^e(\boldsymbol{r}) \tag{61}$$

Eq. (60) is essentially a system of 9 linear equations, and in principle we can solve them to find the 9 components of $v_{ij}^e(\boldsymbol{r})$ as a function of $\tau_i(\boldsymbol{r})$. However, the determinant formed by the numerical coefficients of $v_{ij}^e$ turns out to be zero, so there is no unique solution for each $v_{ij}^e$. Fortunately, we just need to know $v_j^e(\boldsymbol{r})$ as a function of $\tau_i(\boldsymbol{r})$ in order to connect $\gamma_i(\boldsymbol{r})$ with $\boldsymbol{Y}(\boldsymbol{r})$ and $\boldsymbol{v^e}(\boldsymbol{r})$, and $v_j^e(\boldsymbol{r})$ as a function of $\tau_i(\boldsymbol{r})$ can be uniquely decided. Take the example of $v_x^e$, we find that:

$$\begin{aligned} &\partial_x \tau_1 + \partial_x \tau_2 + \partial_x \tau_3 + \partial_y \tau_1 - \partial_y \tau_3 + \partial_z \tau_1 - \partial_z \tau_2 \\ &\qquad = \left(v_{1x}^e + v_{1y}^e + v_{1z}^e\right) + (v_{1x}^e + v_{2x}^e + v_{3x}^e) \end{aligned} \tag{62}$$

Since $v_{ij}^e$ is symmetric, and $v_x^e = v_{1x}^e + v_{2x}^e + v_{3x}^e$, we can immediately solve $v_x^e$ as a function of $\tau_i$. Using similar procedures, we can also solve $v_y^e$ and $v_z^e$, and express all three of them in the following important equation:

$$\begin{aligned} v_x^e &= \frac{1}{2}\left[\partial_x(\tau_1 + \tau_2 + \tau_3) + \partial_y(\tau_1 - \tau_3) + \partial_z(\tau_1 - \tau_2)\right] \\ v_y^e &= \frac{1}{2}\left[\partial_y(\tau_1 + \tau_2 + \tau_3) + \partial_z(\tau_2 - \tau_1) + \partial_x(\tau_2 - \tau_3)\right] \\ v_z^e &= \frac{1}{2}\left[\partial_z(\tau_1 + \tau_2 + \tau_3) + \partial_x(\tau_3 - \tau_2) + \partial_y(\tau_3 - \tau_1)\right] \end{aligned} \tag{63}$$

We thus have established the escape velocity field $\boldsymbol{v}^e(\boldsymbol{r})$ as a function of the regularized metric functions $\tau_1(\boldsymbol{r})$, $\tau_2(\boldsymbol{r})$ and $\tau_3(\boldsymbol{r})$; or equivalently the connection $\boldsymbol{Y}(\boldsymbol{r})$ as a function of the quantum metric functions $\gamma_i(\boldsymbol{r})$.

Put the three components of $\boldsymbol{v}^e$ together, we find that:

$$\boldsymbol{v}^e(\boldsymbol{r}) = \frac{1}{2}\nabla(\tau_1 + \tau_2 + \tau_3) + \frac{1}{2}\left[\partial_y(\tau_1 - \tau_3) + \partial_z(\tau_1 - \tau_2)\right]\vec{\boldsymbol{i}} + \frac{1}{2}[\partial_z(\tau_2 - \tau_1) + \partial_x(\tau_2 - \tau_3)]\vec{\boldsymbol{j}} + \frac{1}{2}[\partial_x(\tau_3 - \tau_2) + \partial_y(\tau_3 - \tau_1)]\,\vec{\boldsymbol{k}} \tag{64}$$

with $\vec{\boldsymbol{i}}, \vec{\boldsymbol{j}},$ and $\vec{\boldsymbol{k}}$ representing the unit vectors along the $x, y, z$ directions, respectively. We denote the first term on the right side of Eq. (64) as $\boldsymbol{v}^t(\boldsymbol{r})$ so

$$\boldsymbol{v}^t(\boldsymbol{r}) = \frac{1}{2}\nabla[\tau_1(\boldsymbol{r}) + \tau_2(\boldsymbol{r}) + \tau_3(\boldsymbol{r})] = \frac{1}{2}\nabla\,\tau(\boldsymbol{r}) \tag{65}$$

Under the local modulus transformation, we find that $\boldsymbol{v}^t(\boldsymbol{r})$ is responsible for the entire change of $\boldsymbol{v}^e(\boldsymbol{r})$ in Eq. (33), while each of the other three terms on the right side of Eq. (64) stays invariant. We denote the sum of these 3 terms as $\boldsymbol{v}^s(\boldsymbol{r})$:

$$\boldsymbol{v}^s(\boldsymbol{r}) = \frac{1}{2}\left[\partial_y(\tau_1 - \tau_3) + \partial_z(\tau_1 - \tau_2)\right]\vec{\boldsymbol{i}} + \frac{1}{2}[\partial_z(\tau_2 - \tau_1) + \partial_x(\tau_2 - \tau_3)]\vec{\boldsymbol{j}} + \frac{1}{2}[\partial_x(\tau_3 - \tau_2) + \partial_y(\tau_3 - \tau_1)]\,\vec{\boldsymbol{k}} \tag{66}$$

so

$$\boldsymbol{v}^e(\boldsymbol{r}) = \boldsymbol{v}^s(\boldsymbol{r}) + \boldsymbol{v}^t(\boldsymbol{r}) \tag{67}$$

If $\boldsymbol{v}^e(\boldsymbol{r})$ is independent of particle mass $m$, then Eq. (63) implies that all 3 $\tau_i$ are also independent of $m$. This can sometimes bring much clarity to our discussion, particularly when gravity, not quantum particle, is the only thing under investigation. In these cases $\tau_i$ will be the preferred metric function to use. On the other hand, when the quantum particle is the focus of study, then $\gamma_i$ is often the preferred choice.

## F. Field equation

We now construct the field equation that relates the external gravitational source's mass density $\rho_{ext}(\boldsymbol{r})$ to the regularized metric functions $\tau_i(\boldsymbol{r})$.

There are 2 basic requirements for the field equation. Namely, it should be covariant to local modulus transformation, and it should match experiments and observations in its applicable domain, *i.e.,* gravity not strong enough to necessitate consideration of relativity-related effects. Let us first use Eq. (32) to replace the gravitational potential energy with its corresponding escape velocity and rewrite the Poisson equation of Eq. (2) as:

$$\boldsymbol{\nabla}^2[\boldsymbol{v}^{\boldsymbol{e}}(\boldsymbol{r})^2] = -8\pi G\rho_{ext}(\boldsymbol{r}) \tag{68}$$

We note that mass density, including that of any external gravitational source, is always invariant to local modulus transformation, so the right side of Eq. (68) is an invariant. However, the left side is dependent on $\boldsymbol{v}^{\boldsymbol{e}}(\boldsymbol{r})$, and $\boldsymbol{v}^{\boldsymbol{e}}(\boldsymbol{r})$ is not invariant to local modulus transformation, so Eq. (68) violates the first requirement for the field equation.

As we already discussed above, $\boldsymbol{v}^{\boldsymbol{e}}(\boldsymbol{r})$ has two parts, $\boldsymbol{v}^{\boldsymbol{s}}(\boldsymbol{r})$ and $\boldsymbol{v}^{\boldsymbol{t}}(\boldsymbol{r})$. While $\boldsymbol{v}^{\boldsymbol{t}}(\boldsymbol{r})$ transforms the same way as $\boldsymbol{v}^{\boldsymbol{e}}(\boldsymbol{r})$, $\boldsymbol{v}^{\boldsymbol{s}}(\boldsymbol{r})$ is invariant to local modulus transformation. A natural solution to fix the left side of Eq. (68) is to replace $\boldsymbol{v}^{\boldsymbol{e}}(\boldsymbol{r})$ by $\boldsymbol{v}^{\boldsymbol{s}}(\boldsymbol{r})$, and we therefore propose the new field equation as:

$$\boldsymbol{\nabla}^2[\boldsymbol{v}^{\boldsymbol{s}}(\boldsymbol{r})^2] = -8\pi G\rho_{ext}(\boldsymbol{r}) \tag{69}$$

In this new equation, both sides are invariant to local modulus transformation.

Let us now use Eq. (66) to replace $\boldsymbol{v}^{\boldsymbol{s}}(\boldsymbol{r})$ by $\tau_i(\boldsymbol{r})$, and we can express the field equation as:

$$\nabla^2\{[\partial_y(\tau_1-\tau_3)+\partial_z(\tau_1-\tau_2)]^2+[\partial_z(\tau_2-\tau_1)+\partial_x(\tau_2-\tau_3)]^2 + [\partial_x(\tau_3-\tau_2)+\partial_y(\tau_3-\tau_1)]^2\} = -32\pi G\rho_{ext}(\boldsymbol{r}) \tag{70}$$

We also know from Eq. (56) that

$$\partial_j\,\tau_i(\boldsymbol{r}) = \partial_i\,\tau_j(\boldsymbol{r}) \tag{71}$$

We note that the particle mass *m* does not appear in Eq. (70), which means $\tau_i(\boldsymbol{r})$ are not dependent on *m*. Eq. (63), in turn, shows that $\boldsymbol{v}^e(\boldsymbol{r})$ is not dependent on *m* either. Thus, we have corroborated the weak equivalence principle in our theory. Furthermore, $\hbar$ does not appear in Eq. (63) and (69), $\boldsymbol{v}^e(\boldsymbol{r})$ is thus not quantum mechanical.

We also note that Eq. (70) is only related to the differences between the $\tau_i$, namely $\tau_1-\tau_2$, $\tau_2-\tau_3$, and $\tau_1-\tau_3$, not their absolute values, and among the 3 differences between $\tau_i$, only 2 are independent. Eq. (70) is only one equation and cannot be expected to solve 2 independent scalar functions. Nevertheless, combining Eq. (70) with Eq. (71) it is plausible that the differences between $\tau_i$ can be solved, and as a result $\boldsymbol{v}^s(\boldsymbol{r})$ can be found. However, $\boldsymbol{v}^s(\boldsymbol{r})$ is only part of $\boldsymbol{v}^e(\boldsymbol{r})$. Since the gravitational potential energy inside the equation of motion Eq. (31) is $-\frac{1}{2}m\boldsymbol{v}^e(\boldsymbol{r})^2$, not $-\frac{1}{2}m\boldsymbol{v}^s(\boldsymbol{r})^2$, one still has the other part $\boldsymbol{v}^t(\boldsymbol{r})$ to consider.

We will discuss $\boldsymbol{v}^t(\boldsymbol{r})$ in detail in the next part. Here, we will just make a sketch of it based on the circumstantial evidence we have. We start by comparing $\boldsymbol{v}^t(\boldsymbol{r})$ with $\boldsymbol{v}^s(\boldsymbol{r})$. First, the magnitude of $\boldsymbol{v}^s(\boldsymbol{r})$ is solely dependent on $\rho_{ext}(\boldsymbol{r})$, and Eq. (69) shows that it is the same as the escape velocity of Newtonian gravity. (Yet, as a vector $\boldsymbol{v}^s(\boldsymbol{r})$ has a well-defined direction, which is quite different from the Newtonian escape velocity.) On the other hand, our discussion above does not provide any link between $\boldsymbol{v}^t(\boldsymbol{r})$ and $\rho_{ext}(\boldsymbol{r})$. Secondly, while $\boldsymbol{v}^s(\boldsymbol{r})$ is dependent on the differences between $\tau_i$, $\boldsymbol{v}^t(\boldsymbol{r})$ depends on their sum. Eq. (65) shows that $\boldsymbol{v}^{\boldsymbol{t}}(\boldsymbol{r})$ is just half of the gradient of $\tau(\boldsymbol{r})$. Thus, $\boldsymbol{v}^{s}(\boldsymbol{r})$ is determined by two independent scalar functions while $\boldsymbol{v}^{\boldsymbol{t}}(\boldsymbol{r})$ by just one scalar function.

At this point, it is essential to clarify the role of local modulus symmetry in the quantum version of our theory versus its classical limit. In the transition from classical mechanics to conventional quantum mechanics, the potential energy is generally kept the same. There is no reason to change that in our theory here. Thus, the potential energy in the classical limit of our theory is $-\frac{1}{2}m\boldsymbol{v}^{\boldsymbol{e}}(\boldsymbol{r})^2$, and the gravitational acceleration is just $\boldsymbol{\nabla}[\frac{1}{2}\boldsymbol{v}^{e}(\boldsymbol{r})^2]$; neither is invariant to local modulus transformation. On the other hand, the discussions in the quantum kinematics and dynamics sections clearly show that all observables are invariant to local modulus transformation in the quantum version of our theory; how can one explain this contradiction? The simple answer to this question is that one must accept a fixed setting of $\tau(\boldsymbol{r})$ and $\boldsymbol{v}^{\boldsymbol{t}}(\boldsymbol{r})$ in the classical limit of our theory. However, this does not violate any fundamental principle, as the concepts of modulus and wave function do not exist in a classical theory. It is true that in both QED and classical electrodynamics, the electromagnetic (EM) field is invariant to local gauge transformation, and the concept of the phase of a matter wave does not exist in classical electrodynamics either. But the EM field is unique due to its exterior derivative nature, while the gravitational acceleration here is just the gradient of gravitational potential energy, thus lacking this unique mathematical property of the EM field. (We also note that in classical electrodynamics, gauge invariance was not considered an essential feature. It only became essential in QFT, particularly for the descriptions of strong and weak interactions.[12]) In the end, if there are no physical and mathematical principles behind a symmetry in a theory, then enforcing such a symmetry is not mandatory in this theory; such is the case for the local modulus symmetry in the classical version of our theory here.

On the other hand, if we start from classical theory, with a given set of $\tau(\boldsymbol{r})$ and $\boldsymbol{v}^{\boldsymbol{t}}(\boldsymbol{r})$, then transition into quantum theory, we can put this same set of $\tau(\boldsymbol{r})$ and $\boldsymbol{v}^{\boldsymbol{t}}(\boldsymbol{r})$ into the wave function and equation of motion. All physical observables should stay the same before and after performing any local modulus transformations, just as described in previous sections.

How does $\boldsymbol{v}^{\boldsymbol{t}}(\boldsymbol{r})$ compare with $\boldsymbol{v}^{s}(\boldsymbol{r})$ in magnitude? Here, the second requirement for the field equation (namely, it should match well with observations) comes into play. When the gravitational field (in our theory with potential energy as $-\frac{1}{2}m\boldsymbol{v}^{\boldsymbol{e}}(\boldsymbol{r})^2$) is well described by Newtonian gravity (with potential energy as $-\frac{1}{2}m\boldsymbol{v}^{s}(\boldsymbol{r})^2$), the second requirement implies that the magnitude of $\boldsymbol{v}^{\boldsymbol{t}}(\boldsymbol{r})$ should be negligibly small compared to that of $\boldsymbol{v}^{s}(\boldsymbol{r})$ in such circumstances. Conversely, in situations where $\boldsymbol{v}^{\boldsymbol{t}}(\boldsymbol{r})$ is not small compared to $\boldsymbol{v}^{s}(\boldsymbol{r})$, one would expect that gravity deviates from Newtonian descriptions.

In summary, $\boldsymbol{v}^{\boldsymbol{t}}(\boldsymbol{r})$ is an irrotational velocity field, with a non-quantum mechanical origin. It is also likely independent of gravitational source mass density. When gravity is Newtonian, $\boldsymbol{v}^{\boldsymbol{t}}(\boldsymbol{r})$ should be negligibly small. In the next part, we will present our idea of the origin of $\boldsymbol{v}^{\boldsymbol{t}}(\boldsymbol{r})$ and quantify it. We then put it into our theory, derive the consequences, and compare them with experiments and observations.

## III. Consequences in an expanding universe

We have mentioned before the similarities between our theory and fluid dynamics and pictured quantum particles moving in a space filled with a flow of purely imaginary velocity field $i\boldsymbol{v}^e(\boldsymbol{r})$. Implicitly presumed in this picture is a static space serving as a fixed background for the flow. What happens when the universe is expanding and the background is not fixed? The second question is, if this flow is entirely caused by gravity, based on the field equation in Eq. (69), why is only part of the flow, *i.e.*, $\boldsymbol{v}^s(\boldsymbol{r})$, related to the gravitational source? Both questions can be answered if we assume that $i\boldsymbol{v}^s(\boldsymbol{r})$ describes the flow in a static space while $i\boldsymbol{v}^e(\boldsymbol{r})$ describes the flow in our expanding universe. Then, $\boldsymbol{v}^s(\boldsymbol{r})$ is completely dependent on the gravitational source and $\boldsymbol{v}^t(\boldsymbol{r})$ is entirely due to the universe's expansion.

Taking any arbitrary point in space as the origin, we can set up a spherical coordinate system of $(r, \theta, \varphi)$. In an expanding universe, at distance $r$ away, space is expanding with a recessionary velocity of $Hr\hat{\boldsymbol{r}}$ relative to the origin, with $H$ as the Hubble parameter and $\hat{\boldsymbol{r}}$ representing the unit vector along the $r$ direction. Let us assume that $i\boldsymbol{v}^s(\boldsymbol{r})$ describes the flow in a static space caused by a gravitational source. When the space is expanding with a recessionary velocity of $Hr\hat{\boldsymbol{r}}$ relative to the origin, we postulate that $i\boldsymbol{v}^e(\boldsymbol{r})$, the flow's velocity field relative to this expanding space, takes the form of $i[\boldsymbol{v}^s(\boldsymbol{r}) - Hr\hat{\boldsymbol{r}}]$ in the spherical coordinate system. This postulate can be equivalently expressed as:

$$\boldsymbol{v}^t(\boldsymbol{r}) = -Hr\hat{\boldsymbol{r}} \tag{72}$$

Furthermore, we postulate that Eq. (72) is true regardless of whether the spatial region under consideration is empty or occupied by any bounded system. Note that $\boldsymbol{v}^t(\boldsymbol{r})$ in Eq. (72) conforms with all the descriptions laid out in the previous subsection.

At this point, it is important to clarify the relationship between our theory and GR in their respective descriptions of the universe's expansion and its consequences in gravity.

In the framework of GR, when relativistic effects can be ignored, gravitationally bound systems like galaxies and stars have essentially Newtonian gravity. Under GR, the universe's expansion plays no role in such systems. Indeed, since objects move in geodesic lines, if the metric for a gravitationally bounded system is of the Robertson-Walker type, gravitationally bounded objects will drift apart over time. This is certainly not consistent with observations.

On the other hand, we argue that the universe's expansion is not intrinsically relativistic. The expansion is quantified by the Hubble parameter $H$ without any involvement of the speed of light. A non-relativistic theory like ours that incorporates the universe's expansion with the expression of Eq. (72) does not violate any fundamental physical principle. Furthermore, both $i\boldsymbol{v}^{t}(\boldsymbol{r})$ and $i\boldsymbol{v}^{s}(\boldsymbol{r})$ are imaginary velocity fields, and their influence on the object's movement is indirect. Quite different from the metric in GR, the expression of $\boldsymbol{v}^{t}(\boldsymbol{r})$ in Eq. (72), which is independent of time $t$, does not lead to gravitationally bounded objects drifting apart over time.

What is the relationship between our theory and GR in their respective descriptions of gravity? Figure 1 suggests that Newtonian gravity is both the non-relativistic limit of GR and the classical limit of our theory. However, the discussion in the previous subsection shows that a finite $\boldsymbol{v}^{t}(\boldsymbol{r})$ causes deviation from Newtonian gravity, even in the non-relativistic and classical limits. If true, this means that even the classical limit of our theory can have new features not predicted by Newtonian gravity and GR, thus implying that GR is not the complete theory of classical gravity. We note that this divergence from GR poses no self-consistency problem here since our theory is outside of the framework of GR, as shown in Fig. 1. On the other hand, cosmology, including the universe's expansion on the cosmological scale, is undoubtedly the exclusive domain of GR. Our theory is non-relativistic and obviously cannot be applied to cosmology. The same is true when gravity is strong, the relativistic effect is not negligible, and GR provides the correct model.

The above discussion suggests that it is in gravitationally bound systems inside an expanding universe that the classical limit of our theory will deviate from the description of GR. It will be shown in the following section that in systems bounded by other forces (mainly the EM force), our theory will have different predictions from conventional quantum mechanics. These results should not come as a surprise. As discussed in the introduction, the theory we try to construct here is ultimately outside the frameworks of GR and conventional quantum mechanics. In the end, if the fundamental physical principles and self-consistency are observed, which theory is correct can only be judged by their comparisons to observations and experiments in their applicable domains. In the following sections, we will explore the consequences of Eq. (72) for the quantum measurement problem, for gravity in gravitationally bound systems, and for that peculiar energy term in Eq. (31).

### A. Quantum measurement problem

Eq. (65) establishes the relationship between $\boldsymbol{v}^t(\boldsymbol{r})$ and $\tau(\boldsymbol{r})$. Based on Eq. (65) and Eq. (72) one finds:

$$\tau(r) = -Hr^2 + \tau_0 \tag{73}$$

with $\tau_0 = \tau(r=0)$. On the other hand, based on Eq. (58), one has:

$$\gamma(r) = e^{\frac{m}{\hbar}\tau(r)} \tag{74}$$

Combining Eq. (73) and (74), one has:

$$\gamma(r) = \gamma_0 \, e^{-\frac{H}{\hbar}mr^2} \tag{75}$$

with $\gamma_0 = \gamma(r = 0)$. Note that since $\tau(\boldsymbol{r})$ is only related to $\boldsymbol{v^t}(\boldsymbol{r})$, $\gamma(r)$ takes the form of Eq. (75) only due to the uniform expansion of the universe. Gravity, represented by gravitational constant *G,* is not involved in Eq. (75).

As discussed in the theoretical part, our theory brought two essential changes to quantum kinematics. Besides the covariant derivative introduced to account for gravity, the other new feature is the co-wave function $\tilde{\psi}(\boldsymbol{r}, t)$. It replaces $\psi^*(\boldsymbol{r}, t)$, and their ratio is the factor $\gamma(\boldsymbol{r})$. Since the expectation value of every physical observable involves both $\psi(\boldsymbol{r}, t)$ and $\tilde{\psi}(\boldsymbol{r}, t)$, $\gamma(r)$, as shown in Eq. (75), has a ubiquitous presence in the new quantum kinematics. We are particularly interested in the changes of $\gamma(r)$ over measurable length scales, as this is the most essential modification of conventional quantum mechanics and most relevant to experiments and observations.

The Gaussian function in Eq. (75) defines a typical length scale $l_m = \sqrt{\frac{\hbar}{Hm}}$, with *m* as the mass of the quantum particle. We call $l_m$ the quantum metric length. Much below $l_m$, the change brought by $\gamma(r)$ is negligible. Above $l_m$, the quantum particle's co-wave function and its probability density function decay exponentially as any Gaussian function. Note that since $\hbar$ and *H* are both constants, $l_m$ is only dependent on *m,* with $l_m \propto m^{-1/2}$. In other words, *m* is the only changeable parameter of the Gaussian function $\gamma(r)$.

Taking the value of the current Hubble constant $H_0$ at $2.27 \times 10^{-18} s^{-1}$, we find that for electrons, the lightest and most important microscopic particles in our everyday life, $l_m \sim 7 \times 10^7 m$. Look at it in another way, over $1\, cm$, $\gamma(r)$ changes by about $2 \times 10^{-18}$. Thus, for electrons, the effect of $\gamma(r)$ can be safely ignored. On the other extreme, take the example of a macroscopic object of typical density, with a mass of 1 $kg$ and a size of about 10 $cm$. In this case, one finds $l_m \sim 7 \times 10^{-8} m$, or about $10^{-6}$ of its actual size. In such a case, any relevant length scale of the object is so much greater than $l_m$ that $\gamma(r)$ can be taken as a Dirac delta function in all practical applications.

The two examples above show the binary nature of our theory's experimental predictions. For microscopic objects, in general, the change of $\gamma(r)$ over any relevant laboratory scale is too small to be detected. Thus, our theory produces practically the same results for microscopic objects as conventional QM. On the other hand, for any macroscopic object, the drastic change of $\gamma(r)$ renders the object's co-wave function $\tilde{\psi}(\boldsymbol{r}, t)$ to a quantum mechanical pointer state over any relevant length scale for such an object. This is independent of how its wave function $\psi(\boldsymbol{r}, t)$ evolves dynamically in the Schrödinger equation. Instead, this new feature $\gamma(r)$ in quantum kinematics can cause the "collapse" of the probability density function in real space without the "collapse" of the wave function. Consequently, for macroscopic objects, quantum spatial fluctuation is not possible kinematically for any length much more than $l_m$.

This binary result between microscopic and macroscopic objects offers a straightforward solution to the quantum measurement problem. The discussion above shows that position basis vectors are the preferred basis vectors for incorporating gravity and the universe's expansion into quantum mechanics. For microscopic objects, since $\gamma(r)$ is practically constant over space, this requirement is not essential, so any other basis can be used. For macroscopic objects, with $\gamma(r)$ being practically a Dirac delta function, position basis becomes the only choice. Thus, the preferred-basis problem[13] has a natural solution in our theory.

To address the problem of definite outcomes[13], for simplicity, we assume that a microscopic object initially is in a superposition ket state of $|\varphi\rangle = c_1|\varphi_1\rangle + c_2|\varphi_2\rangle$, and the measuring macroscopic apparatus is in a neutral ket state of $|\Psi_0\rangle$. We note that $|\varphi\rangle$ can be of any basis while $|\Psi\rangle$ can only be of position basis, as discussed above. Suppose $|\varphi_1\rangle$ generating the measuring outcome of $|\Psi_1\rangle$ and $|\varphi_2\rangle$ of $|\Psi_2\rangle$. Like the standard quantum mechanics, we also assume the time evolution of the combined system's state is linear. Thus, after the microscopic superposition state $|\varphi\rangle = c_1|\varphi_1\rangle + c_2|\varphi_2\rangle$ is measured, the combined system is in a final ket state of $|f\rangle = c_1|\varphi_1\rangle \otimes |\Psi_1\rangle + c_2|\varphi_2\rangle \otimes |\Psi_2\rangle$.

Let us assume the spatial separation between any two adjacent pointer states is some experimentally feasible length $l_p$. Since $l_m \ll l_p$ for macroscopic objects, our theory shows that the effect of $\gamma(r)$ on $\tilde{\psi}(\boldsymbol{r},t)$ makes it kinematically impossible to have a superposition of pointer bra states for the measuring apparatus. (We note that bra vector and co-wave function are treated as equivalent here). On the other hand, we can take any point in space as the origin of our coordinate system. This means that $\gamma(r)$ can be centered on any position, which should result in no bias between pointer states $|\Psi_1\rangle$ and $|\Psi_2\rangle$. Thus, after measurement, one can have either $\langle\Psi_1|$ or $\langle\Psi_2|$ for the apparatus. For the combined system, its final bra state $\langle f|$ is either $c_1\langle\varphi_1| \otimes \langle\Psi_1|$ or $c_2\langle\varphi_2| \otimes \langle\Psi_2|$. Note that $\langle\varphi_1|\varphi_2\rangle = \langle\Psi_1|\Psi_2\rangle = 0$, and all physical quantities after measurement are decided by both $\langle f|$ and $|f\rangle$. Thus, each time after the measurement is performed, only one pointer state, *i.e.,* one definite outcome of the measuring apparatus, appears, with the probability of $|c_1|^2$ or $|c_2|^2$, the same as predicted from the amplitude of the microscopic superposition state. The generalization of this argument to any amplitude distribution of microscopic eigenstates is straightforward, hence the solution to the quantum measurement problem.

As shown in Eq. (75), the exponent of $\gamma(r)$ is proportional to the object's mass. With a change of more than 20 orders of magnitude between the masses of microscopic and macroscopic objects, it is easy to see why their $\gamma(r)$ show such a binary outcome over the laboratory length scale, *i.e.*, either being flat or being a delta function. It is also for this reason that we call $\tau(\boldsymbol{r})$ the regularized metric function, as the drastic change of $\gamma(\boldsymbol{r})$ as a function of mass *m* is stripped away from $\tau(\boldsymbol{r})$, besides $\tau(\boldsymbol{r})$ being a logarithmic function of $\gamma(\boldsymbol{r})$. If we again take 1 $g/cm^3$ as the mass density for a typical object, based on Eq. (75), one can find the mass of an object that has its quantum metric length $l_m$ the same as its physical size. This mass turns out to be about $4 \times 10^{-5}$ $g$, or about twice of the Planck mass. Evidently, this mass value depends on the mass density $\rho$ one uses. However, since the size of an object scales as $\rho^{-1/3}$, we expect the order of magnitude of this mass value to stay the same for most liquid and solid objects. As a result, one can answer Bell's question of where the dividing line between microscopic and macroscopic objects is [14], and it is the object's mass that matters. If an object's mass is much more than $10^{-5}$ $g$, such an object can be called macroscopic and behave classically; for objects with mass much below $10^{-5}$ $g$, they can be called microscopic, and conventional quantum mechanics can describe their behaviors adequately. Thus, from the perspective of our theory, objects with a mass of around $10^{-5}$ $g$ are the most interesting, as they are expected to show behavior that is neither classical nor the same as conventional QM describes. Of course, to detect the change of $\gamma(r)$ over a length of $l_m \sim 7 \times 10^{-4} m$, one needs to maintain $\psi(\boldsymbol{r}, t)$, *i.e.*, quantum coherence over such a distance for an object with a mass of about $10^{-5}$ $g$. That is indeed challenging experimentally.

Penrose has long advocated that gravity causes the nearly total disappearance of quantum spatial fluctuation for macroscopic objects.[15] Our theory argues that the ultimate cause is the universe's expansion. Based on Eq. (75), it is the Hubble constant that determines the dividing mass line. On the other hand, our theory shows that this dividing mass line is only a practical one, and it is worth stressing here that in our theory, a unified physical law governs both microscopic and macroscopic objects. Indeed, it is this unified physical law that predicts this dividing line. The Copenhagen interpretation instead divides the world into micro and macro worlds with different rules for each.[16] In light of the binary outcome for macroscopic and microscopic objects implied in Eq. (75), this division is probably a necessary arrangement in the early development of quantum mechanics. Nevertheless, it is a practical and temporary solution. From the perspective of our theory, to justify a temporary solution by insisting on the impossibility of a unifying law for both macroscopic and microscopic objects even in principle is completely unnecessary.

A related topic is the tension between the asymmetry of time in the second law of thermodynamics and the time-reversal symmetry of dynamic law in classical mechanics first and quantum mechanics later.[15] The fundamental dynamic law of QM has time-reversal symmetry. If one makes the transformation of $t \to -t$, $\psi^*(\boldsymbol{r}, -t)$ is still the solution of the complex conjugate of the Schrödinger equation, provided that the potential energy of the system is real. Based on Eq. (31), this is still the case in our theory, so $\psi^*(\boldsymbol{r}, -t)$ is still the solution. However, the co-wave function of the system, after time reversal transformation, is $\tilde{\psi}(\boldsymbol{r}, -t)$, not $\psi^*(\boldsymbol{r}, -t)$. As long as $\gamma(\boldsymbol{r})$ is not constant, the two are not the same. In principle, $\psi^*(\boldsymbol{r}, -t)$ does not have any physical meaning for the quantum system in our theory, while $\tilde{\psi}(\boldsymbol{r}, -t)$ is not a solution of the complex conjugate of the dynamic equation. Thus, time-reversal symmetry is broken at the molecular level, at least in principle, even though the spatial change of $\gamma(\boldsymbol{r})$ for microscopic objects is very small, as we have shown above. Is this time-reversal symmetry breaking at the molecular level responsible for the asymmetry of time in the second law of thermodynamics? This is an open and important question that demands more investigation.

### B. Gravitationally bounded systems

In this section, we focus on gravity and examine quantitatively how gravity departs from the Newtonian description in an expanding universe. For simplicity, we consider the case where the gravitational source is a mass point located at the origin of the spherical coordinate system, with a mass *M*. For a test particle located at any point $\boldsymbol{r}$**,** the unit mass potential energy is $u_g(\boldsymbol{r}) = -\frac{1}{2}\boldsymbol{v}^e(\boldsymbol{r})^2 = -\frac{1}{2}\boldsymbol{v}^s(\boldsymbol{r})^2 - \frac{1}{2}\boldsymbol{v}^t(\boldsymbol{r})^2 - \boldsymbol{v}^s \cdot \boldsymbol{v}^t$.

Based on the field equation in Eq. (69), the first term is the Newtonian gravitational potential energy, and in our case, one has:

$$-\frac{1}{2}\boldsymbol{v}^s(\boldsymbol{r})^2 = -\frac{GM}{r} \qquad (76)$$

Based on Eq. (72), the second term $-\frac{1}{2}\boldsymbol{v}^t(\boldsymbol{r})^2 = -\frac{1}{2}H^2r^2$. Notice that this term is independent of *M*, *i.e*., it exists even without gravitational source mass. Since we define $r$ as the distance between the source and the test particle, without a gravitational source, $-\frac{1}{2}H^2r^2$ is not meaningful so far as the gravity between the source mass and the test particle is concerned. Therefore, we will treat it as a background energy irrelevant to our discussion here. The third term, $-\boldsymbol{v}^s \cdot \boldsymbol{v}^t$, is the coupling between $i\boldsymbol{v}^s$ and $i\boldsymbol{v}^t$. Since $\boldsymbol{v}^s$ is dependent on *M*, this term is also dependent on *M*. Therefore, its gravitational effect should be considered. Indeed, it is this term that causes gravity in our theory to depart from Newtonian gravity. Unlike the first term, which only depends on the magnitude of $\boldsymbol{v}^s$, one needs to know the vector $\boldsymbol{v}^s$ to calculate $-\boldsymbol{v}^s \cdot \boldsymbol{v}^t$.

From Eq. (66), we know that $\boldsymbol{v}^s$ is decided by the first-order spatial derivative of $\tau_i$. In a spherical coordinate system, $\tau_i$ with $i = 1, 2, 3$ represent the 3 regularized metric functions in the $\hat{\boldsymbol{r}}, \widehat{\boldsymbol{\theta}}, \widehat{\boldsymbol{\varphi}}$ directions, respectively. We also express the gradient of each $\tau_i$ as $\boldsymbol{\nabla}\tau_i = \partial_r \tau_i \, \hat{\boldsymbol{r}} + \partial_\theta \tau_i \, \widehat{\boldsymbol{\theta}} + \partial_\varphi \tau_i \, \widehat{\boldsymbol{\varphi}}$, with the spatial derivatives in the spherical coordinate defined as $\partial_r = \frac{\partial}{\partial r}$, $\partial_\theta = \frac{1}{r}\frac{\partial}{\partial \theta}$, $\partial_\varphi = \frac{1}{r \sin\theta}\frac{\partial}{\partial \varphi}$. In the following calculation we will not need these exact expressions of $\partial_j$ $(j = r, \theta, \varphi)$ because we do not need to solve $\tau_i$. Our goal is to solve $\boldsymbol{v}^s$, so finding out $\partial_j \tau_i$ is enough.

Combining Eq. (65) with Eq. (72), one has 3 conditions for $\partial_j \tau_i$:

$$\begin{aligned} &\partial_r \tau_1 + \partial_r \tau_2 + \partial_r \tau_3 = -\,2Hr \\ &\partial_\theta \tau_1 + \partial_\theta \tau_2 + \partial_\theta \tau_3 = 0 \\ &\partial_\varphi \tau_1 + \partial_\varphi \tau_2 + \partial_\varphi \tau_3 = 0 \end{aligned} \tag{77}$$

Eq. (71) shows $\partial_j \tau_i(\boldsymbol{r}) = \partial_i \tau_j(\boldsymbol{r})$, which amounts to 3 more conditions for $\partial_j \tau_i$. We also note that since the gravitational source is at the origin, this case has spherical symmetry. Thus, for each $\tau_i$ one expects:

$$\partial_\theta \tau_i = \partial_\varphi \tau_i \tag{78}$$

The magnitude of $\boldsymbol{v}^s$ is $\sqrt{\frac{2GM}{r}}$, the same as the Newtonian escape velocity. Combining this condition in Eq. (76) with Eq. (71), (77), and (78), we have 9 independent algebraic equations for $\partial_j \tau_i$, which can be solved easily. Put the solutions of $\partial_j \tau_i$ into Eq. (66), one finds two solutions for $\boldsymbol{v}^s$**:**

$$\boldsymbol{v}^s = \pm\sqrt{\frac{GM}{3r}}\left(2\hat{\boldsymbol{r}} - \hat{\boldsymbol{\theta}} - \hat{\boldsymbol{\varphi}}\right) \tag{79}$$

Combining the solution of $\boldsymbol{v}^t$ in Eq. (72) and $\boldsymbol{v}^s$ in Eq. (79) and discarding the background $-\frac{1}{2}\boldsymbol{v}^t(\boldsymbol{r})^2$ term, one has $u_g(\boldsymbol{r}) = -\frac{GM}{r} \pm \sqrt{\frac{4GMr}{3}}H$. Since the gravitational acceleration $\boldsymbol{a}_g = -\boldsymbol{\nabla} u_g$, there are also two solutions for $\boldsymbol{a}_g$. For gravitationally bound systems, stronger attraction should be the favored solution during the formation of these systems. Consequently, we will only use the stronger acceleration solution in the following discussions. It is expressed as:

$$\boldsymbol{a}_g = -\left(\frac{GM}{r^2} + \sqrt{\frac{GM}{3r}}H\right)\hat{\boldsymbol{r}} \tag{80}$$

The first term in Eq. (80), not surprisingly, is the Newtonian inverse square term. The second coupling term is inversely proportional to the square root of the distance and proportional to the Hubble parameter.

To compare these two terms in some real-world examples, we first take the Earth as the gravitational source and approximate Earth's mass distribution as a mass point located at the center of the Earth. Again, we set the Hubble constant at its current value of $2.27 \times 10^{-18} s^{-1}$. We find that on the Earth's surface, the coupling term is about $10^{-14}\, ms^{-2}$, or about $10^{-15}$ of Newtonian gravity, and well below any current experimental limit. In the second example, we take the sun as the gravitational source. The sun's Newtonian gravity on Earth is about $6 \times 10^{-3}\, ms^{-2}$, the coupling term is about $4 \times 10^{-14}\, ms^{-2}$ on Earth, and the ratio between them is about $10^{-11}$. The coupling term is still completely negligible, but the ratio increases by about $10^4$ from the first example.

There are two things worth noting here. First, due to the small value of the Hubble constant, the gravitational effect caused by the coupling term is very small and negligible in the Earth and solar system. This is as expected; otherwise, its effect should have already been detected. Since the solar system is a typical stellar system, we expect the coupling term's effect to be negligible in other stellar systems as well. Second, the coupling term decreases as the square root of the distance, much slower than the Newtonian gravity. As a result, the coupling term is expected to play a more important role at larger distances.

Starting from the observation by Zwicky in the 1930s,[17], it is noticed that the dynamics of galaxies and galaxy clusters often cannot be adequately described by Newtonian gravity if one uses the observed baryonic mass of these galaxies and galaxy clusters as the sole gravitational source. ΛCDM, the current standard model of cosmology, attributes such deviation, or "mass discrepancy", to the existence of nonbaryonic dark matter inside galaxies and galaxy clusters.[18] The process of Big Bang nucleosynthesis and large-scale structure formation evolved from the initial condition written on the cosmic microwave background also provide strong argument for dark matter. Yet, despite many efforts, no direct detection of dark matter has ever been confirmed.

Could the coupling term in Eq. (80) provide an alternative explanation for the "mass discrepancy" in the galactic systems? The ratio between the coupling and the Newtonian terms in Eq. (80) is $\sqrt{\frac{H^2}{3GM}}r^{3/2}$. Thus, when the distance is larger than $\sqrt[3]{\frac{3GM}{H^2}}$, denoted here as the critical length $l_c$, the coupling term is expected to dominate over the Newtonian term, and gravity is expected to become distinctively stronger compared to Newtonian prediction. Based on Eq. (80), stars that are circularly orbiting around the centers of galaxies should have their rotational velocity as follows:

$$v = \sqrt{\frac{GM}{r} + \sqrt{\frac{GMr}{3}}H} \qquad (81)$$

Here again we take the crude approximation that the mass *M* of the galaxy is concentrated at its center point. Thus, when the star's distance to the center of the galaxy *r* is well below $l_c$, the rotation curve is expected to show the Newtonian behavior of $v \approx \sqrt{\frac{GM}{r}}$, *i.e.*, $v$ decreases with $r$ as $v \propto r^{-1/2}$ ; when $r \gg l_c$, one expects $v \approx (\frac{GH^2}{3})^{\frac{1}{4}} M^{\frac{1}{4}} r^{\frac{1}{4}}$ , *i.e.*, now $v$ increases very slowly with *r* as $v \propto r^{1/4}$. Equally important, $v \propto M^{1/4}$ in this range.

To better illustrate the effect of the coupling term in the galactic dynamics, we study a hypothetical spiral galaxy. We model the bulge of the galaxy as a sphere with a radius *R* of 3 kpc and a density $\rho$ of $6 \times 10^{-21} kg\, m^{-3}$, and we assume all the galaxy's mass, *i.e.*,$1.99 \times 10^{40} kg$ or $10^{10}$ of the solar mass, is uniformly distributed in the spherical bulge. Inside the bulge, only the mass that is within $r$ has net contribution to gravity, as a result $\frac{r^3}{R^3} M$ should replace *M* in Eq. (81). Outside the bulge, no change is made to Eq. (81).

The Newtonian prediction is a linear increase of the rotational velocity with increasing distance inside the bulge, followed by a decrease of $v \propto r^{-\frac{1}{2}}$ outside of the bulge. The result is plotted in the dashed line in Fig. 2 above. At the edge of the bulge where $r = R$, compared to the Newtonian result, Eq. (81) predicts the rotational velocity increases by a factor of $\sqrt{1 + \sqrt{\frac{H^2}{4\pi G\rho}}}$ where $\rho$ is the mass density of the bulge. Using the current Hubble constant $H_0$ and mass density above, we find ${H_0}^2$ is less than $4\pi G\rho$ by almost 6 orders of magnitude. As a result, one has to go further out to $r \sim 100R$ to find a significant increase of rotational velocity compared to the Newtonian result. This is not consistent with observations of most spiral galaxies.[19]

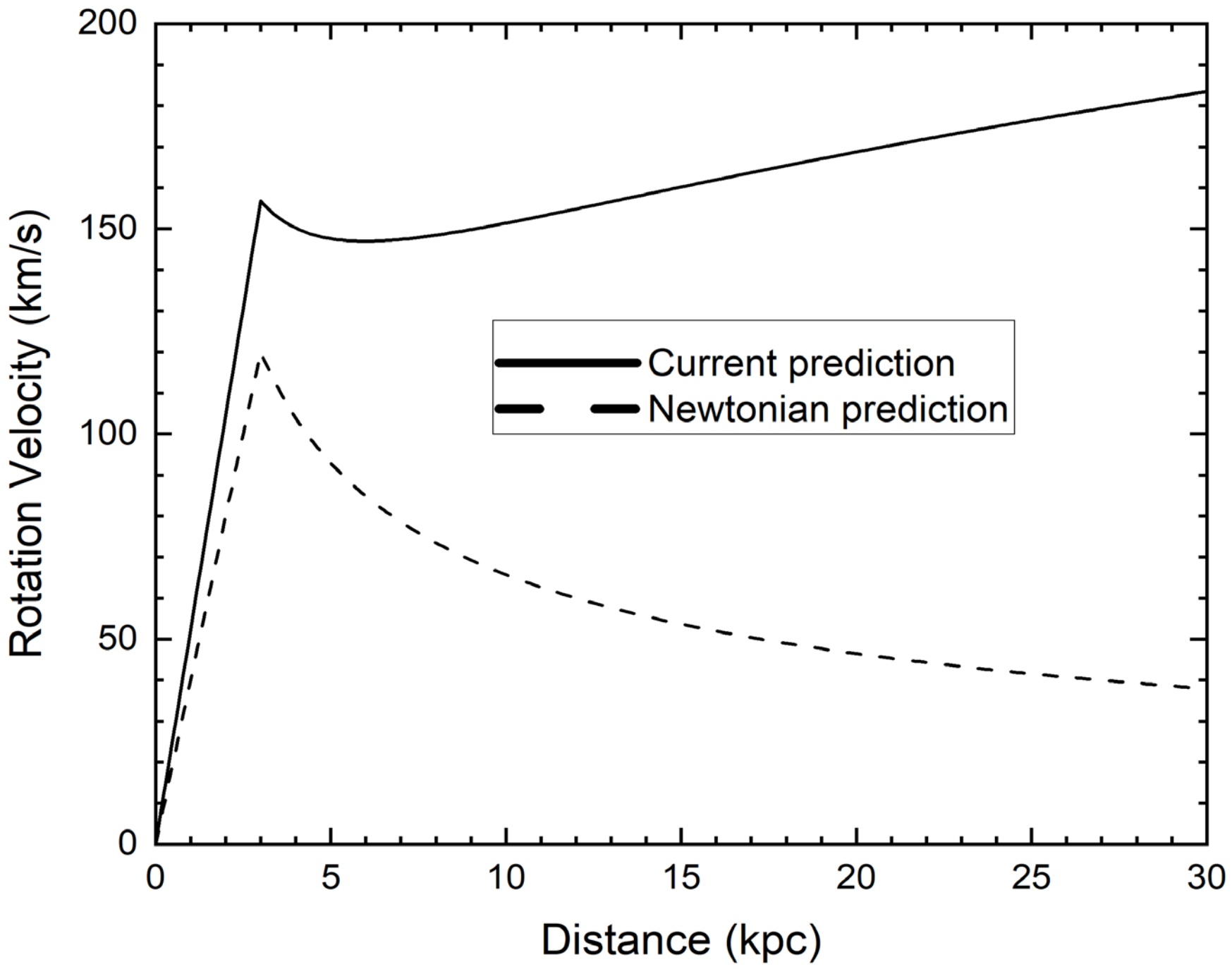

*Figure 2: The rotational velocity (in km/s) of stars in a hypothetical spiral galaxy as a function of the distance (in kpc) between the center of the galaxy and the stars. For simplicity, the bulge of the galaxy is modeled as a sphere with a radius of 3 kpc and a uniform density of* $6 \times 10^{-21} kg\ m^{-3}$*, and all the galaxy's mass is concentrated in the bulge. The result is plotted in a solid line, with the Hubble parameter set at* $H = 700H_0 = 1.59 \times 10^{-15} s^{-1}$. *The Newtonian prediction is plotted in the dashed line.*

On the other hand, if the Hubble parameter *H* takes a value close to $\sqrt{4\pi G\rho}$, one expects gravity to significantly deviate from Newtonian descriptions already at the edge of the galactic bulge. In Fig. 2 above, we have set $H = 700H_0 = 1.59 \times 10^{-15} s^{-1}$ and plotted the result in a solid line based on Eq. (81) (with $M$ replaced by $\frac{r^3}{R^3}M$ for $r \leq R$). The resemblance to a typical spiral galaxy's rotation curve is evident.**Error! Bookmark not defined.** The rotational velocity also increases linearly inside the bulge, with a slope about 30% steeper than the Newtonian result. Outside of the bulge, the velocity first decreases slightly, then increases slowly with distance. The initial decrease is not an inherent feature, as a higher value of $H$ can eliminate it.

The calculations above are based on approximations of uniform density inside the galaxy bulge and no mass outside it. The reality is of course more complicated. There can be significant mass outside of the bulge and $\rho$ is not uniform inside it. Any detailed comparison with observation must consider the mass distribution function. Also, in following the formula of Eq. (81), we have assumed that $\boldsymbol{v}^t$ is still $-Hr\hat{\boldsymbol{r}}$, even though the gravitational source is no longer a point source. Again, this is an approximation. Nevertheless, under such crude approximations, the result, as illustrated by the solid line of Fig. 2, can capture the main features of the observed rotation curve of a typical spiral galaxy. Moreover, as pointed out earlier, $v \propto M^{1/4}$ in the range of $r \gg l_c$, which is the well-known Tully-Fisher relation.[20] This makes us tentatively believe that our theory can potentially provide a mechanism to account for the "mass discrepancy" in the galactic systems.

However, one key element still needs explanation, *i.e.*, why is there such a big difference between the Hubble parameter we use ($700H_0$) and the current Hubble constant $H_0$? At the time of the initial formation of galaxies, the universe's expansion rate was much higher than now. One possible explanation is that at the structural formation stage of the galaxies, the Hubble parameter at that time was, to use our example, $700H_0$, and gravity is much stronger than the Newtonian description. Once galactical structures were formed, this Hubble parameter was somehow "frozen" in the $\boldsymbol{v^t}$ of the bounded systems. Afterwards, even as the universe's expansion rate decreased substantially, $\boldsymbol{v^t}$ does not change in these galactic systems. As a result, one needs to set $H = 700H_0$ instead of the current rate of $H_0$ in Eq. (81) to describe the rotation curve of our hypothetical spiral galaxy.

If what we described above was at work throughout the universe's history, many understandings based on ΛCDM will need rethinking, including when and how the galaxies formed. However, before one can claim the scenario outlined above is feasible, one needs to answer the two essential questions: Can a "frozen" mechanism in the $\boldsymbol{v^t}$ be found during the formation of galaxies? Was the Hubble parameter with such a high value at the time of galactic structural formation consistent with a cosmological model that incorporates our theory? Clearly, much more investigation needs to be done, particularly on structural formation, before we can draw any concrete conclusions.

There are other alternative theories for the mass discrepancy problem, notably MOND, first proposed by Milgrom.[21] MOND was partially built on the basis of the Tully-Fisher relation, so it is not surprising that our theory is in agreement with MOND on this aspect. However, MOND predicts flat rotational curves at large distances, while our theory predicts $v \propto r^{1/4}$ at large distances. MOND uses the light-mass ratio as the adjusting parameter to fit observed rotation curves; we can use the Hubble parameter as the adjusting parameter after taking into account the mass distribution function of the galaxies.

The key feature in MOND is the characteristic and universal acceleration $a_0$, estimated at $1.2 \times 10^{-10} ms^{-2}$. From Eq. (80) we know that when the Newtonian acceleration $\frac{GM}{r^2} = \frac{H^2 r}{3}$, the coupling term contributes the same amount of acceleration as the Newtonian term. In other words, this is where acceleration starts to deviate significantly from Newtonian prediction. Using the edge of the bulge as the reference point where the rotation velocity starts to change with distance very slowly, we set $r = R = 3\ kpc$, and $H$ is still set at $700H_0$. Putting these parameters into Eq. (80) generates an acceleration of $1.56\ \times 10^{-10} ms^{-2}$. The discussion above is meant only as a rough estimate, so a 30% difference from the MOND acceleration should be considered quite good. On the other hand, to maintain a constant $a_0$ across galaxies with very different masses and sizes, one would require $H^2R$ as a constant in our theory for these galaxies. Again, the studies of galaxy formation are imperative to verify whether that is true or not.

In the end, MOND is a phenomenological theory, while our theory is a fundamental theory that traces the origin of the deviation from Newtonian gravity to the universe's expansion. Much more studies are needed to see whether our theory can produce the phenomenological success of MOND and whether it can remedy the weaknesses of MOND, particularly in the description of galaxy groups and clusters.[22]

Note that even though our theory is a theory of quantum mechanics and gravity, the modification of Newtonian gravity discussed in this section is entirely non-quantum mechanical and non-relativistic. In some way, this transformation is similar to the one from electrostatics to electrodynamics. In E&M, one may start with the Coulomb force law in electrostatics, with the ultimate coupling strength being the electron charge *e*. Electrostatics then can be transformed into electrodynamics based on the relativistic principle, with the involvement of universal constant *c* and the modification caused by it. Here, one starts with Newtonian gravity, which has the universal constant *G* as the coupling strength. (Indeed, the Coulomb force law and the Newtonian gravitational law are very similar.) This Newtonian force law is subsequently modified by the effect of the universe's expansion quantified by the Hubble parameter *H*. Between these two transformations, there is a reverse of order in the introduction of universal constants *G* and *c* versus parameters *H* and *e*. At the same time, this also draws attention to the analogy between *H* and *e*. The difference is that while *e* is a running coupling constant that changes over space and momentum but not over time, *H* changes over time but is a constant over space.

### C. the energy term

In the theoretical part, we briefly mentioned that the term $-\frac{\hbar}{2}\boldsymbol{\nabla}\cdot\boldsymbol{v^e}(\boldsymbol{r})$ inside the equation of motion of Eq. (31) looks quite peculiar; let us now examine it more closely. The potential energy of a test particle with mass *m* in the Newtonian gravitational field of a point source with mass *M* is $-\frac{GMm}{r}$, *i.e.,* it is proportional to both *M* and *m*. We have mentioned the unit mass energy $-\frac{1}{2}H^2r^2$ before, which becomes $-\frac{1}{2}mH^2r^2$ for test particle with mass *m*. Since it is unrelated to the source mass *M*, we have put it into the background and ignored it in the discussions of gravity between the source and the test particle. On the other hand, the term $-\frac{\hbar}{2}\boldsymbol{\nabla}\cdot\boldsymbol{v^e}(\boldsymbol{r})$, as a potential energy for the test particle in Eq. (31), is not related to its mass *m*. This suggests that it should also be put into the background and ignored in the discussions of gravity between the source and the test particle.

We now decompose $-\frac{\hbar}{2}\nabla\cdot\boldsymbol{v}^e(\boldsymbol{r})$ into $-\frac{\hbar}{2}\nabla\cdot\boldsymbol{v}^s(\boldsymbol{r})-\frac{\hbar}{2}\nabla\cdot\boldsymbol{v}^t(\boldsymbol{r})$. The first term is dependent on *M* but not on *m*, as discussed above, the second term is even more peculiar. Using Eq. (72), we can express it as:

$$-\frac{\hbar}{2}\nabla\cdot\boldsymbol{v}^t(\boldsymbol{r}) \; = \; \frac{1}{2}\hbar H \tag{82}$$

This means it depends neither on the source nor the test particle; it is a purely background positive energy. It is quantum mechanical due to the $\hbar$ in it, and ubiquitous as it is proportional to *H*. This brings up the possibility of its association with dark energy.

Dark energy is thought necessary to explain the universe's accelerating expansion.[23] The current estimate puts this energy density at around $6\times10^{-10}Jm^{-3}$ throughout the universe.[24] The vacuum energy in QFT is far greater than the observed value to account for its origin. Using Planck energy (about $2\times10^{9}J$) divided by the Planck volume (about $4\times10^{-105}m^3$) as an approximation, the vacuum energy density is estimated at around $5\times10^{113}Jm^{-3}$, or about 123 orders of magnitude higher.

Set the Hubble parameter *H* at its current value $H_0$ again, we find that $\frac{1}{2}\hbar H$ is about $10^{-52}$ Joule. To compare it with QFT's vacuum energy density and the current estimate of dark energy density, we divide $\frac{1}{2}\hbar H_0$ by the Planck volume, which translates into an energy density at about $2.5\times10^{52}Jm^{-3}$. This is about 61 orders of magnitude larger than the current estimate of the dark energy density and about 61 orders of magnitude less than the vacuum energy density, so it is almost exactly halfway between them. Is this pure coincidence, or is there any physical significance in it? At this point, we do not know. There are a few difficult questions to answer before one can make some sense of these numbers. Our theory is non-relativistic, while Planck scale quantities, including Planck energy and volume, contain all 3 universal constants. Is it justified to divide the energy $\frac{1}{2}\hbar H_0$ by the Planck volume to produce an energy density? If not, then what is the alternative? Also, what happens to the vacuum energy density? Even if some other mechanism originates the dark energy, one still needs to answer why the vacuum energy density does not exist. Suppose the Planck energy over time evolves into $\frac{1}{2}\hbar H_0$ and the Planck volume increases by the same factor so that the energy density decreases by the square of this factor. In that case, these questions can have a self-consistent answer and the $\frac{1}{2}\hbar H_0$ energy term can produce an energy density that matches the current estimate of the dark energy density. Then again, we do not know anything about the mechanism for such an evolution, if it exists at all. It seems likely that we will ultimately need the theory for the Planck scale to answer all these questions.

## IV. Relation with other theories

Our theory bears resemblances to quantum field theory. The local modulus symmetry here is mathematically similar to the *U(1)* gauge symmetry in quantum electrodynamics. The resulting covariant derivative in our theory is thus mathematically similar to that in quantum electrodynamics. The classical version of QED is classical electrodynamics. As we discussed above, there are also similarities between the classical version of our theory and classical electrodynamics. Consequently, the Newtonian gravity at the first level of Fig. 1 needs modification due to the universe's expansion, just like electrostatics needs to be transformed into electrodynamics under relativistic principles.

Our theory also shares similarities with GR, and there are two contributing factors. As we point out in the beginning of the theory part, quantum mechanics and special relativity have structural similarities, particularly in their shared use of vector and covector in linear space to describe states. When they are fused with gravity, many of these features remain intact. The second factor is the equivalence principle. Even though the exact statements are not the same, as they refer to the reduction of laws to SR and QM in small enough regions, respectively, the essence is the same for both cases. It is the expression of gravity through kinematics instead of dynamics, as gravity is not a normal force associated with conventional dynamic law, and it can be transformed away locally by certain kinematic transformations. This is the fundamental feature of gravity that other forces do not share. If the four-dimensional spacetime is curved by gravity in GR, in some sense, one can similarly state that the Hilbert space of the quantum system in position representation is curved by gravity in our theory.

Ultimately, the purpose of our theory is to unite quantum mechanics with gravity in a non-relativistic context, just as the purpose of GR is to unite SR with gravity in a non-QM context, while QFT unites SR with QM without gravity. As we discussed in the introduction, in such a framework, they share the same characteristics of encompassing two fundamental constants and remaining invariant to certain local transformations. While the first-level theories in Fig. 1 do not have local symmetry, the requirement of local symmetry seems necessary to fuse two of them together at a time.

While QFT is the theory for the smallest scale, and GR is the theory for the largest, it seems the theory we propose here occupies a scale in between. For small particles and molecules, the suggested deviation from QM is currently too small to be detected. Consequently, our theory is unlikely to have much implication for elementary particle physics, which is well described by QFT. On the other hand, the kinematics of human-scale objects can be significantly modified by our theory, as discussed above.

On the gravitational side, the suggested modification of Newtonian law becomes evident only when gravity is very weak. This is the opposite to GR, where one expects more deviations from Newtonian law with stronger gravity. Such a separation in gravitational strength between our theory and GR also makes it possible to apply them separately in different situations. On the largest scale, the universe itself, it is clear that GR is the basic foundational theory that underpins cosmology.

The in-between status of the theory proposed here is not limited to the scale of its applications but also features in its theoretical structure. In QED, the connection in the covariant derivative is the 4-potential, while in GR, the connection is the gravitational field. Here, we have a connection $\boldsymbol{Y}(\boldsymbol{r})$, or equivalently, the velocity field $\boldsymbol{v}^{\boldsymbol{e}}(\boldsymbol{r})$, that is neither a potential nor a field. Instead, the potential is $-\frac{1}{2}m\boldsymbol{v}^{e}(\boldsymbol{r})^{2}$. Taking the gradient of this potential yields a field that is a product of the connection $\boldsymbol{v}^{\boldsymbol{e}}(\boldsymbol{r})$ and its spatial derivatives. Thus, the field is neither like the connection in GR nor like the connection's exterior derivative in QED, but somewhere in between as a combination of both. Whether there is any deeper meaning of such a structure awaits further exploration.

At the end of this section, it is natural to ask about the relativistic parent of this theory. By extending the perspective presented in Fig. 1 to one level above, it seems clear that such a relativistic parent theory should also be the gravitational parent theory of QFT and the quantum mechanical parent theory of GR. In other words, it is the theory of ultimate accuracy, with $G$, $\hbar$, and $c^{-1}$ all set at their actual values. Clearly, such a theory is essential for the description of Planck scale physics. However, for most situations in the current universe, a combination of GR, QFT, and potentially the theory proposed here may be sufficient, with different situations calling for different theories or the right combinations of them. This may even be true for most of the universe's evolutionary history, with the notable exceptions of the very early stage of the Big Bang and the center of black holes where Planck scale physics is clearly needed. On the other hand, non-relativistic characteristics are embedded in our theory here. For example, the imaginary momentum field $im\boldsymbol{v}^{\boldsymbol{e}}(\boldsymbol{r})$ is automatically removed from the particle momentum expectation value based on the rules in our theory. This fits well with the fact that a non-relativistic field has no momentum of its own. Changing this to a relativistic version does not look compatible with the structure of our current theory. Thus, the road to a relativistic parent theory does not look like to be an easy extension. Again, this is not very surprising, considering the difficulties people encounter quantizing the gravitational field or fusing QFT and GR together.

## V. Conclusions

Quantum field theory and general relativity, the two most fundamental theories in physics, share the common features of possessing local symmetry and encompassing two universal constants ($\hbar$, $c$ and $G$, $c$, respectively). Inspired by these commonalities, we built a non-relativistic theory encompassing $G$ and $\hbar$ with local symmetry. Starting with the observation that the modulus of the wave function can be changed by an arbitrary constant factor in conventional quantum mechanics, we elevate this global symmetry into a local one. The requirement of local modulus symmetry results in several important changes from conventional quantum mechanics. Foremost is replacing the complex conjugate of wave function by a new entity called co-wave function. They differ by a real and positive factor of $\gamma(\boldsymbol{r})$ that may change from position to position. Another important change is the replacement of the ordinary spatial derivative by a covariant spatial derivative, with a purely imaginary connection $im\boldsymbol{v}^{\boldsymbol{e}}(\boldsymbol{r})$ attached to the canonical momentum operator to form a particle momentum operator. The ordinary temporal derivative also needs to be replaced by a covariant temporal derivative resembling the Lagrangian derivative in fluid dynamics, with a purely imaginary velocity field of $i\boldsymbol{v}^{\boldsymbol{e}}(\boldsymbol{r})$. With these changes, we can make the modified Schrödinger equation covariant to local modulus transformation. We find the potential energy term now is $-\frac{1}{2}m\boldsymbol{v}^{\boldsymbol{e}}(\boldsymbol{r})^2$. This can be identified as gravitational potential energy if $\boldsymbol{v}^{\boldsymbol{e}}(\boldsymbol{r})$ is identified as the escape velocity.

To account for gravity, we define 3 quantum metric functions $\gamma_i(\boldsymbol{r})$, with their product equal to $\gamma(\boldsymbol{r})$, and their corresponding regularized metric functions $\tau_i(\boldsymbol{r})$ are defined too. We postulate that in small enough regions, physical laws reduce to conventional quantum mechanics in our theory. Based on this new equivalence principle, we are able to find out $\boldsymbol{v}^{\boldsymbol{e}}(\boldsymbol{r})$ as a function of $\tau_i(\boldsymbol{r})$. It turns out that $\boldsymbol{v}^{\boldsymbol{e}}(\boldsymbol{r})$ can be divided into two parts, with one part $\boldsymbol{v}^{\boldsymbol{s}}(\boldsymbol{r})$ that is invariant to local modulus transformation and another part $\boldsymbol{v}^{\boldsymbol{t}}(\boldsymbol{r})$ that takes up all the changes required for $\boldsymbol{v}^{\boldsymbol{e}}(\boldsymbol{r})$ under local modulus transformation. Replacing $\boldsymbol{v}^{\boldsymbol{e}}(\boldsymbol{r})$ by $\boldsymbol{v}^{\boldsymbol{s}}(\boldsymbol{r})$ in the potential energy term of $-\frac{1}{2}m\boldsymbol{v}^{\boldsymbol{e}}(\boldsymbol{r})^2$, we find a new gravitational Poisson equation invariant to local modulus transformation. Finally, we find that $\boldsymbol{v}^{\boldsymbol{t}}(\boldsymbol{r})$ takes the form of $-Hr\hat{\boldsymbol{r}}$ in an expanding universe with $H$ as the Hubble parameter.

With the theory established, we explore its consequences in 3 areas. First, we find that $\gamma(\boldsymbol{r})$ is a Gaussian function of $r$ with an exponent proportional to the particle mass and the Hubble constant. Simple calculations show that $\gamma(\boldsymbol{r})$ barely registers any change for microscopic objects. For macroscopic objects, $\gamma(\boldsymbol{r})$ changes so rapidly with *r* that the co-wave functions become pointer states. This provides a solution to the quantum measurement problem. A related consequence of the changing $\gamma(\boldsymbol{r})$ is the breaking of time reversal symmetry in the quantum system. This may have implications for the time asymmetry in the second law of thermodynamics. Next, we study our theory's effect on gravity in the classical limit. Taking a mass point as the gravitational source, we find that the point source's gravitational acceleration field has an inverse square root term proportional to the Hubble parameter, in addition to the Newtonian inverse square term. Compared to the Newtonian term, gravity due to this new term is negligible up to the stellar systems, then can become the more dominant one on the galactic scale. We calculate the rotation curve of a hypothetical spiral galaxy. We find that it can match observations of real spiral galaxies well only if we set the value of the Hubble parameter to several hundred times the current Hubble constant. Whether the Hubble parameter was "frozen" in the $\boldsymbol{v^t}$ of the gravitationally bounded systems during their formations, and whether such high values of the Hubble parameter can be found at the time when galaxies started to form so that one can justify using it to fit the rotation curve, are still open questions. Lastly, we find the energy term $-\frac{\hbar}{2}\boldsymbol{\nabla}\cdot\boldsymbol{v^t}(\boldsymbol{r})$ in the equation of motion yields a very small energy of $\frac{1}{2}\hbar H_0$ in our expanding universe.

[18] P. J. E. Peebles, *Growth of the nonbaryonic dark matter theory,* Nat Astro **1,** 57 (2017).

[19] R. H. Sanders and S. S. McGaugh, *Modified Newtonian dynamics as an alternative to dark matter*, Ann. Rev. Astron. Astrophys. **40**, 263 (2002).

[20] R. B. Tully and J. R. Fisher, *A new method of determining distances to galaxies*, Astron. Astrophys. **54**, 661 (1977).

[21] M. Milgrom, *A modification of the Newtonian dynamics as a possible alternative to the hidden mass hypothesis,* Astrophys. J. **270**, 371 (1983).

[22] S. McGaugh, *A Tale of Two Paradigms: the Mutual Incommensurability of $\Lambda$CDM and MOND*, Canadian Journal of Physics **93**, 250 (2015).

[23] P. J. E. Peebles and B. Ratra, *The cosmological constant and dark energy,* Review of Modern Physics **75**, 559 (2003).

[24] P. A. R. Ade et al. (Planck Collaboration), *Planck 2013 results. I. Overview of products and scientific results*, Astron. Astrophys. **571**, A1 (2014).